\pdfoutput=1
\newcommand*{\ATLASLATEXPATH}{}
 
\documentclass[PAPER, atlasdraft=true, cernpreprint,  texlive=2018, UKenglish, LANGEDIT=true, LANGSHOW=false]{\ATLASLATEXPATH atlasdoc}
 
\usepackage[backend=biber]{\ATLASLATEXPATH atlaspackage}
\usepackage{\ATLASLATEXPATH atlasbiblatex}
 
\usepackage[process=true]{\ATLASLATEXPATH atlasphysics}
 
\newcommand\T{\rule{0pt}{2.6ex}}       
\newcommand\B{\rule[-1.2ex]{0pt}{0pt}} 
\newcommand{\hsp}{\hspace*{0.3cm}}
 
\usepackage{lscape}
\usepackage{afterpage}
\usepackage{multirow}
\usepackage{rotating}
\usepackage{amssymb} 
 
\usepackage{booktabs}
\usepackage{siunitx}
 
\graphicspath{{logos/}{figures/final/}}
 
\usepackage{ANA-HIGG-2018-04-PAPER-defs}

\AtlasTitle{Observation of $H \rightarrow b\bar{b}$ decays and $VH$ production with the ATLAS detector}

\AtlasAbstract{
A search for the decay of the Standard Model Higgs boson into a $b\bar{b}$ pair when
produced in association with a $W$ or $Z$ boson is performed with the ATLAS detector. The data, corresponding to an integrated luminosity of $79.8\,\mathrm{fb}^{-1}$ were collected in
proton--proton collisions during Run~2 of the Large Hadron Collider at a centre-of-mass energy
of $13\,\text{Te\kern -0.1em V}$. For a Higgs boson mass of $125\,\text{Ge\kern -0.1em V}$, an excess of events over the expected
background from other Standard Model processes is found with an observed (expected)
significance of 4.9 (4.3) standard deviations. A combination with the results
from other searches in Run 1 and in Run 2 for the Higgs boson in the $b\bar{b}$ decay mode is performed, which yields
an observed (expected) significance of 5.4 (5.5) standard deviations, thus
providing direct observation of the Higgs boson decay into $b$-quarks. The ratio of the
measured event yield for a Higgs boson decaying into $b\bar{b}$ to the Standard Model
expectation is $1.01 \pm 0.12 \mathrm{(stat.)} ^{+0.16}_{-0.15} \mathrm{(syst.)}$.
Additionally, a combination of Run 2 results searching for the Higgs boson produced in
association with a vector boson yields an observed (expected) significance of 5.3 (4.8)
standard deviations.
}

\AtlasRefCode{HIGG-2018-04}
 
\PreprintIdNumber{CERN-EP-2018-215}

\arXivId{1808.08238}

\AtlasJournalRef{\PLB 786 (2018) 59}
\AtlasDOI{10.1016/j.physletb.2018.09.013}
 
\AtlasCoverEgroupEditors{atlas-ana-higg-2018-04-contact-editors.cern.ch}
 
\AtlasCoverEgroupEdBoard{atlas-ana-higg-2018-04-editorial-board@cern.ch}
 
\hypersetup{pdftitle={ATLAS document},pdfauthor={The ATLAS Collaboration}}
 
\begin{document}
 
\maketitle
 
\tableofcontents
 
\clearpage

\section{Introduction}
% The next lines are included from the .//paper_01_intro.tex input file
The Higgs boson~\autocite{Englert:1964et,Higgs:1964ia,Higgs:1964pj,Guralnik:1964eu} was
discovered in 2012 by the ATLAS and CMS
Collaborations~\autocite{HIGG-2012-27,CMS-HIG-12-028} with a mass of approximately 125\,\GeV\, from the analysis of
proton--proton ($pp$) collisions produced by the Large Hadron Collider (LHC)~\autocite{Evans:2008zzb}. Since then,
the analysis of data collected at centre-of-mass energies of 7\,\TeV, 8\,\TeV\ and 13\,\TeV\
in Runs 1 and 2 of the LHC has led to the observation of many of the production
modes and decay channels predicted by the Standard Model (SM). The bosonic decay
channels are well established and have entered an era of precision
measurements~\cite{HIGG-2014-14, HIGG-2016-21, HIGG-2016-22, HIGG-2016-25, HIGG-2016-33,
HIGG-2017-11, Aaboud:2018jqu, CMS-HIG-17-025, CMS-HIG-16-040, CMS-HIG-16-041,
CMS-HIG-16-042}. The decay into $\tau$-lepton pairs was first observed in the
combination of the ATLAS and CMS analyses~\cite{HIGG-2015-07,CMS-HIG-16-043}. The main Higgs boson
production modes, gluon--gluon fusion (ggF) and vector-boson fusion (VBF), were already measured following the analysis of the Run~1 data, and
recently the coupling of the Higgs boson to top quarks was directly observed by
the ATLAS and CMS Collaborations~\autocite{HIGG-2018-13, CMS-HIG-17-035} through the observation of
the associated production of a Higgs boson and a top-quark pair ($\ttbar H$).
 
The dominant decay of the SM Higgs boson is into pairs of $b$-quarks, with an expected
branching fraction of approximately 58\% for a mass of $\mh=125$~\GeV~\autocite{Djouadi:1997yw}. However, large
backgrounds from multi-jet production make a search in the dominant gluon--gluon fusion production mode very challenging at hadron colliders.
The most sensitive production modes for detecting \Hbb\ decays are the associated production of a Higgs
boson and a $W$ or $Z$ boson~\cite{Glashow:1978ab} ($VH$), where the leptonic decay of the
vector boson enables efficient triggering and a significant reduction of the multi-jet
background. As well as probing the dominant decay of the Higgs boson, this measurement allows
the overall Higgs boson decay width~\cite{Lafaye:2009vr,Heinemeyer:2013tqa} to be constrained and
provides the best sensitivity to the $ZH$ and $WH$ production modes, which are (for instance) important elements in the interpretation of Higgs boson measurements in effective field
theories~\cite{Ellis:2014dva}.
 
Searches in this channel at the Tevatron by the CDF and D0 Collaborations showed an excess
of events with a significance of 2.8 standard deviations for a Higgs boson with a mass of
125\,\GeV~\cite{TEV-HIGGS-COMB-12}. Analysing the 2015 and 2016 data and combining
with the Run~1 results~\cite{CMS-HIG-13-012, HIGG-2013-23}, both the ATLAS
and CMS Collaborations reported evidence for Higgs boson production and decay in this channel, with
observed (expected) significances of 3.6 (4.0) and 3.8 (3.8) standard deviations,
respectively~\cite{HIGG-2016-29,CMS-HIG-16-044}. Searches for
\Hbb\ decays have also been conducted in the VBF~\cite{HIGG-2014-12, CMS-HIG-14-004, HIGG-2016-30} and
$\ttbar H $~\cite{HIGG-2013-27, HIGG-2015-05, HIGG-2017-03, CMS-HIG-14-010, CMS-HIG-17-026, CMS-HIG-17-022} channels, and with high transverse momentum Higgs bosons~\cite{CMS-HIG-17-010}, but
with markedly lower sensitivities.
 
This Letter reports an update to the search for the SM Higgs boson decaying into a $b\bar{b}$ pair in the
$VH$ production mode with the ATLAS detector in Run 2 of the LHC
presented in Ref.~\cite{HIGG-2016-29}. This update uses \lumiFull\ of $pp$
collision data collected at a centre-of-mass energy of 13\,\TeV, to be compared with 36.1\,\ifb\ for the previous result. In addition, an updated version of the ATLAS reconstruction code and improved object calibrations are used, the impact of the luminosity and modelling systematic uncertainties are reduced from updated measurements and estimates, and larger samples of simulated events are used to model the background processes. Events are selected
in 0-, 1- and 2-lepton channels, based on the number of charged leptons, $\ell$ (electrons or
muons), to explore the $ZH \to \nu\nu b\bar{b}$,  $WH \to \ell \nu b \bar{b}$ and $ZH \to
\ell\ell b\bar{b}$ signatures, respectively. The dominant background processes after the
event selection are \vjets, \ttbar, single-top and diboson process. Multivariate discriminants, built from
variables that describe the kinematics of the selected events, are used to maximise the sensitivity
to the Higgs boson signal. Their output distributions are combined using a binned maximum-likelihood fit, referred to as the global likelihood fit,
which allows the signal yield and the background normalisations to be extracted. The signal extraction method is cross-checked with the dijet-mass analysis, where the signal yield is extracted using the mass of the dijet system as the main fit observable, and
validated using the diboson analysis, where the nominal multivariate analysis is modified to extract the $VZ$, $Z \rightarrow b\bar{b}$ diboson process. The result of the multivariate analysis is then combined
with that of the previously published analysis of Run~1
data~\autocite{HIGG-2013-23}, with other searches for $\bbbar$ decays of the Higgs
boson and with other searches
in the $VH$ production mode.
The latter two combinations lead to the observation of both the $\bbbar$ decay of the Higgs
boson and $VH$ production. An observation of the $\bbbar$ decay of the Higgs boson by the
CMS Collaboration~\cite{Sirunyan:2018kst} was submitted for publication at the same time as this Letter.
% End of text imported from the .//paper_01_intro.tex input file

\section{The ATLAS detector}
% The next lines are included from the .//paper_02_atlas.tex input file
ATLAS~\cite{PERF-2007-01} is a general-purpose particle detector covering nearly the entire solid angle\footnote{ATLAS uses a right-handed coordinate system with its origin at the nominal interaction point (IP) in the centre of the detector and the $z$-axis coinciding with the axis of the beam pipe.  The $x$-axis points from the IP towards the centre of the LHC ring, and the $y$-axis points upward. Cylindrical coordinates ($r$,$\phi$) are used in the transverse plane, $\phi$ being the azimuthal angle around the $z$-axis. The pseudorapidity is defined in terms of the polar angle $\theta$ as $\eta = - \ln \tan(\theta/2)$. The distance in ($\eta$,$\phi$) coordinates, $\Delta R = \sqrt{(\Delta\phi)^2+(\Delta\eta)^2}$, is also used to define cone sizes. Transverse momentum and energy are defined as $\pt=p\sin\theta$ and $\et=E\sin\theta$, respectively.}
around the collision point.  An inner tracking detector, located within a 2~T axial
magnetic field generated by a thin superconducting solenoid, is used to measure the
trajectories and momenta of charged particles. The inner layers consist of
high-granularity silicon pixel detectors covering a pseudorapidity
range $|\eta| < 2.5$, and include an innermost layer~\cite{Capeans:1291633,Abbott:2018ikt} that was added to the detector between Run 1 and Run 2.
Silicon microstrip detectors covering $|\eta| < 2.5$ are located beyond the pixel detectors.
Outside the microstrip detectors and covering $|\eta | < 2.0$, there are straw-tube tracking detectors, which also provide measurements of transition radiation that are used in electron identification.
A calorimeter system surrounds the inner tracking detector, covering the pseudorapidity range $|\eta| < 4.9$.
Within the region $|\eta|< 3.2$, electromagnetic calorimetry is provided by barrel ($|\eta| < 1.475$) and
endcap ($1.375 < |\eta| < 3.2$) high-granularity lead/liquid-argon (LAr) sampling calorimeters,
with an additional thin LAr presampler covering $|\eta| < 1.8$ to correct for energy loss in material upstream of the calorimeters.
Hadronic calorimetry is provided by a steel/scintillator-tile calorimeter within $|\eta| < 1.7$, and copper/LAr endcap calorimeters extend the coverage to $|\eta|=3.2$.
The solid angle coverage for $|\eta|$ between 3.2 and 4.9 is completed with copper/LAr and tungsten/LAr calorimeter modules optimised for electromagnetic and hadronic measurements, respectively.
The outermost part of the detector is the muon spectrometer, which measures the curved trajectories of muons in the magnetic field of three large air-core superconducting toroidal magnets. High-precision tracking is performed within the range $|\eta| < 2.7$ and there are chambers for fast triggering within the range $|\eta| < 2.4$.
A two-level trigger system~\cite{Aaboud:2016leb} is used to reduce the recorded data rate.
The first level is a hardware implementation aiming to reduce the rate to around
100~kHz, while the software-based high-level trigger provides the remaining rate reduction to approximately 1~kHz.

% End of text imported from the .//paper_02_atlas.tex input file
 
\section{Object and event selection}
% The next lines are included from the .//paper_04_selection.tex input file
The event topologies characteristic of $VH$,~$H\to\bbbar$ processes considered contain zero, one or
two charged leptons, and two `$b$-jets' containing particles from $b$-hadron decays. The
object and event selections follow those of
Ref.~\cite{HIGG-2016-29} to a large extent.
 
\subsection{Object reconstruction}
 
Tracks measured in the inner detector are used to reconstruct interaction
vertices~\cite{Aaboud:2016rmg}, of which the one with the highest sum of squared
transverse momenta of associated tracks is selected as the primary vertex.
 
Electrons are reconstructed from topological clusters of energy deposits in the
calorimeter~\cite{ATL-PHYS-PUB-2017-022} and matched to a track in the
inner detector. Following Refs.~\cite{ATLAS-CONF-2016-024,HIGG-2016-29}, \textit{loose} electrons are required
to have $\pt>7\,\GeV$ and $\abseta<2.47$, to have small impact
parameters,\footnote{Transverse and longitudinal impact parameters are defined relative
to the primary vertex position, where the beam line is used to approximate the primary
vertex position in the transverse plane.} to fulfil a loose track isolation requirement,
and to meet a `LooseLH' quality criterion computed from shower shape and
track quality variables~\cite{Aaboud:2016vfy}. In the 1-lepton channel, \textit{tight}
electrons are selected using a `TightLH' likelihood requirement and a stricter
calorimeter-based isolation.
 
Muons are required to be within the acceptance of the muon spectrometer $\abseta<2.7$, to
have $\pt>7\,\GeV$, and to have small impact parameters. \textit{Loose} muons are selected
using a `loose' quality criterion~\cite{PERF-2015-10} and a loose track
isolation. In the 1-lepton channel, \textit{tight} muons fulfil the `medium' quality
criterion and a stricter track isolation.
 
Hadronically decaying $\tau$-leptons~\autocite{PERF-2014-06,ATLAS-CONF-2017-029} are
required to have $\pt>20\,\GeV$ and $\abseta<2.5$, to be outside of the transition region
between the barrel and end-cap electromagnetic calorimeters $1.37<\abseta<1.52$, and to meet a `medium' quality
criterion~\cite{ATLAS-CONF-2017-029}. They are only used in the analysis to avoid $\tau$-leptons being misidentified as jets.
 
Jets are reconstructed from topological clusters~\cite{ATL-LARG-PUB-2008-002}
using the \antikt\ algorithm~\cite{Cacciari:2008gp} with radius
parameter $R=0.4$. A jet vertex tagger~\cite{Aad:2015ina} is used to remove jets
associated with vertices other than the primary one
for jet $\pt<60\,\GeV$ and $\abseta<2.4$. Jet cleaning criteria
are used to identify jets arising from non-collision backgrounds or noise in the
calorimeters~\cite{ATLAS-CONF-2015-029} and events containing such jets
are removed. Jets are required to have $\pt>20\,\GeV$ in the central region
($\abseta<2.5$), and $\pt>30\,\GeV$ outside ($2.5<\abseta<4.5$) of the tracker acceptance.
In the central region, they are tagged as containing $b$-hadrons using a multivariate discriminant~\autocite{ATL-PHYS-PUB-2017-013} (MV2), with the selection tuned to produce an average efficiency
of 70\% for $b$-jets in simulated $\ttbar$ events, which corresponds to light-flavour ($u$-, $d$-, $s$-quark and gluon) and $c$-jet misidentification efficiencies of 0.3\% and 12.5\% respectively.
 
Simulated jets are labelled as $b$-, $c$- or light-flavour jets according to which hadrons with
$\pt>5\,\GeV$ are found within a cone of size $\Delta R=0.3$ around their
axis.
Simulated $V$+jets events are categorised depending on the labels of the jets that form
the Higgs boson candidate: $V+ll$ when they are both light-flavour jets, $V+cl$ when there is one
$c$-jet and one light-flavour jet, and $V$+HF (heavy flavour) in all other cases, mainly two $b$-jets. Owing to
the large rejection of light-flavour  jets achieved by the MV2 discriminant, simulated $V+ll$,
$V+cl$ and $WW$ events are not subjected to the $b$-tagging requirement due to the resulting low number of simulated events, but instead they are weighted by the probability that their jets pass the $b$-tagging selection~\cite{HIGG-2016-29}.
 
In addition to the standard jet energy scale
calibration~\autocite{PERF-2016-04}, $b$-tagged jets receive additional
flavour-specific corrections to improve their energy measurement (scale and resolution): if any muons are found within $\Delta
R=0.4$, the four-momentum of the closest muon is added to that of the jet, and a residual correction is applied to equalise the
response to jets with leptonic or hadronic decays of heavy-flavour hadrons. In the
2-lepton channel, a per-event kinematic likelihood uses the full reconstruction of the
event kinematics to improve the estimate of the energy of the $b$-jets. The corrections
improve the resolution of the dijet mass by up to 40\%~\cite{HIGG-2016-29}.
 
The missing transverse momentum $\vecmet$ is reconstructed as the negative vector sum of the
momenta of leptons, hadronically decaying $\tau$-leptons and jets, and of a `soft term' built from additional tracks matched
to the primary vertex~\autocite{PERF-2016-07}. The
magnitude of $\vecmet$ is referred to as \met.
An overlap removal procedure is applied to avoid any double-counting between the
reconstructed leptons, including the hadronically decaying $\tau$-leptons, and jets.

\subsection{Event selection and categorisation}
 
Events are categorised into the 0-, 1- and 2-lepton channels depending on the number of
selected electrons and muons, to target the $ZH \to vv b\bar{b}$, $WH \to \ell \nu b \bar{b}$
and $ZH \to \ell\ell b\bar{b}$ signatures, respectively. In all channels, events are
required to have exactly two $b$-tagged jets, which form the Higgs boson candidate. At least one
$b$-tagged jet is required to have \pt\ greater than 45\,\GeV. Events are further split into 2-jet or
3-jet categories depending on whether additional, untagged jets are present. In the 0- and 1-lepton
channels, only one such jet is allowed, as the \ttbar\ background is much larger in events
with four jets or more. In the 2-lepton channel any number of jets
is accepted in the 3-jet category.
 
The reconstructed transverse momentum \ptv\ of the vector boson corresponds to \met\ in the 0-lepton
channel, to the vectorial sum of $\vecmet$ and the charged-lepton transverse momentum in the
1-lepton channel, and to the transverse momentum of the 2-lepton system in the 2-lepton
channel. As the signal-to-background ratio increases for large Higgs boson transverse
momenta~\autocite{Butterworth:2008iy,ATL-PHYS-PUB-2009-088},
the analysis focuses on a high-\ptv\ region defined as $\ptv>150\,\GeV$. In the 2-lepton
channel, the sensitivity is increased by the addition of a medium-\ptv\ region
with $75\,\GeV<\ptv<150\,\GeV$.
 
Two versions of the analysis are carried out, one using a multivariate approach and the other using the dijet mass as the final discriminant. The event selection shown in Table~\ref{tbl:evselTable} is applied to both versions, with further selections applied for the dijet-mass analysis. The two versions of the analysis also have different event categorisations, with further details outlined below.
 
% The next lines are included from the .//tables/table_04_evtsel.tex input file
\afterpage{
\begin{landscape}
\begin{table}[ptbh]
\centering
\caption{Summary of the event selection and categorisation in the 0-, 1- and 2-lepton channels.}
\label{tbl:evselTable}
\begin{tabular}{l c c c c}
\toprule
\multirow{2}{*}{Selection} & 0-lepton & \multicolumn{2}{c}{1-lepton} & 2-lepton\\
&                & $e$ sub-channel & $\mu$ sub-channel &   \\
\midrule
Trigger & \met\ & Single lepton & \met\  & Single lepton \B \\
\multirow{2}{*}{Leptons} & 0 \textit{loose} leptons & 1 \textit{tight} electron & 1 \textit{tight} muon  & 2 \textit{loose} leptons with $\pt>7$~\GeV \\
&   with $\pt>7$~\GeV                    & $\pt>27$~\GeV  & $\pt>25$~\GeV   & $\ge$ 1 lepton with $\pt>27$~\GeV \\
\met\ & $>$ 150~\textrm{\GeV} & $>$ 30~\textrm{\GeV} & -- & --  \\
\mll\ & -- & \multicolumn{2}{c}{--}  & 81~\textrm{\GeV}~$<$ \mll\ $<$ 101~\textrm{\GeV} \\
\midrule
Jets  & \multicolumn{3}{c}{Exactly 2 / Exactly 3 jets} & Exactly 2 / $\geq$ 3 jets \\
\multirow{2}{*}{Jet \pt} &  \multicolumn{4}{c}{$>$ 20 \textrm{\GeV} for $|\eta|$~$<$~2.5} \\
& \multicolumn{4}{c}{$>$ 30 \textrm{\GeV} for 2.5~$<$~$|\eta|$~$<$~4.5} \\
$b$-jets & \multicolumn{4}{c}{Exactly 2 $b$-tagged jets} \\
Leading $b$-tagged jet \pt\  & \multicolumn{4}{c}{$>$ 45 \textrm{\GeV}} \\
\midrule
$H_{\mathrm{T}}$ & $>$ 120 \textrm{\GeV} (2 jets), $>$150 \textrm{\GeV} (3 jets) & \multicolumn{2}{c}{--}  & -- \\
$\textrm{min}[\Delta\phi(\vecmet,\vec{\mathrm{jets}}) ]$ & $> 20\ensuremath{^\circ}$ (2 jets), $> 30\ensuremath{^\circ}$ (3 jets) & \multicolumn{2}{c}{--}  & -- \\
$\Delta\phi(\vecmet, \vec{bb})$ & $> 120\ensuremath{^\circ}$ & \multicolumn{2}{c}{--}  & -- \\
$\Delta\phi(\vec{b_1},\vec{b_2})$ & $< 140\ensuremath{^\circ}$ & \multicolumn{2}{c}{--}  & -- \\
$\Delta\phi(\vecmet,\vec{p}_{\mathrm{T}}^{\mathrm{miss}})$ & $< 90\ensuremath{^\circ}$ & \multicolumn{2}{c}{--}  & -- \\
\midrule
$\pt^V$ regions & \multicolumn{3}{c}{$>150$~\GeV} & 75~\textrm{\GeV}~$<$ $\pt^V$ $<$ 150~\textrm{\GeV}, $>150$~\GeV \T \B \\
\midrule
\multirow{2}{*}{Signal regions} & \multirow{2}{*}{--} &
\multicolumn{2}{c}{\multirow{2}{*}{$\mbb \geq 75$~\GeV\ or $m_{\mathrm{top}}\leq225$~\GeV}} & Same-flavour leptons \\
&                   &            \multicolumn{2}{c}{}         & Opposite-sign charges ($\mu\mu$ sub-channel)\\
\midrule
\multirow{2}{*}{Control regions} & \multirow{2}{*}{--}
&\multicolumn{2}{c}{\multirow{2}{*}{$\mbb < 75$~\GeV\ and
$m_{\mathrm{top}}>225$~\GeV}} & Different-flavour
leptons\\
&                   &            \multicolumn{2}{c}{}         & Opposite-sign charges \\
\bottomrule
\end{tabular}
\end{table}
\end{landscape}
}
% End of text imported from the .//tables/table_04_evtsel.tex input file
 
\paragraph{0-lepton channel} The online selection uses \met\ triggers with
thresholds that varied from 70\,\GeV\ to 110\,\GeV\ between the 2015
and 2017 data-taking periods. Their efficiency was measured in $W$+jets, $Z$+jets and $\ttbar$ events
in data using single-muon triggers, resulting in correction factors that are applied to
the simulated events, ranging from 1.05 at the offline \met\ threshold of 150\,\GeV\ to a
negligible deviation from unity at an \met\ above 200\,\GeV.
A requirement on the scalar sum of the transverse momenta $H_\mathrm{T}$ of the jets removes a
small part of the phase space where the trigger efficiency depends mildly on the number of
jets in the event. Events with any \textit{loose} lepton are rejected. High \met\ in multi-jet
events typically arises from mismeasured jets in the calorimeters. Such events are
efficiently removed by requirements on the angular
separation of the $\vecmet$, jets, and $\vec{p}_{\mathrm{T}}^{\mathrm{miss}}$ (the
missing transverse momentum calculated using only tracks reconstructed in the inner tracking detector and
matched to the primary vertex).
 
\paragraph{1-lepton channel} In the electron sub-channel, events are required to satisfy a logical
OR of single-electron triggers with identification and isolation criteria looser than
those used in the offline analysis, and \pt\ thresholds that started at 24\,\GeV\ in 2015
and increased to 26\,\GeV\ in 2016 and 2017. The muon sub-channel uses the same \met\
triggers and correction factors as the 0-lepton channel, as these triggers effectively
select on \ptv\ given that muons are not included in the online \met\ calculation and they perform more efficiently than the single-muon triggers in the analysis phase
space. Events are required to have exactly one high-\pt\ \textit{tight} electron or muon,
and no additional \textit{loose} leptons. In the electron sub-channel an additional
selection of $\met>30\,\GeV$ is applied to reduce the background from multi-jet
production. Events are categorised into the signal region (SR) or into a control region
enriched in $W$+HF events ($W$+HF CR) using selections on the invariant mass of the two
$b$-tagged jets ($m_{bb}$), and on the reconstructed mass of a semi-leptonically
decaying top-quark candidate ($m_{\mathrm{top}}$). The latter is calculated as the invariant mass of
the lepton, the reconstructed neutrino\footnote{The transverse component of the neutrino
momentum is identified with $\vecmet$, and the longitudinal component is obtained
by constraining the lepton--neutrino system to the $W$ mass.}
and the $b$-tagged jet that yields the lowest mass value. The resulting purity of
the $W$+HF control region is around 75\%.

\paragraph{2-lepton channel} The online selection in the electron sub-channel is the same
as in the 1-lepton channel. In the muon sub-channel, a similar OR of single-muon triggers
is used, with lowest \pt\ thresholds increasing with luminosity and ranging from 20\,\GeV\ to 26\,\GeV.
Events must have exactly two \textit{loose} leptons, one of which must have
$\pt>27\,\GeV$, and the invariant mass of the lepton pair must be compatible with that of
the $Z$ boson. Events with same-flavour leptons
enter the signal region, while events with one muon and one electron define an
$e\mu$ control region which is over 99\% pure in $\ttbar$ and single-top-quark events.
 
The acceptances in the three channels after the event selection, as well as the predicted
cross-sections times branching fractions for \wzh\ with \wln, \zll, \znn, and $H\to\bbbar$
are given in Table~\ref{tab:crosssection_acc}. The non-negligible acceptance for the $qq\to
WH$ process in the 0-lepton channel is mostly due to events with an unidentified hadronically decaying
$\tau$-lepton produced in the $W$ decay, while the larger acceptance for the $gg\to ZH$
process compared with $qq\to ZH$  is due to the harder \ptv\ spectrum of the gluon-induced
process.
% The next lines are included from the .//paper_04b_acceptance.tex input file
\begin{table}[tb!]
\begin{center}
\caption{The cross-section ($\sigma$) times branching fraction ($\mathcal{B}$) and acceptance for the three channels
at $\sqrt{s}=13$~\TeV.
The $qq$- and $gg$-initiated $ZH$ processes are shown separately.
The branching fractions are calculated considering only decays into muons and electrons for \zll,
decays into all three lepton flavours for \wln\ and decays into all neutrino flavours for \znn.
The acceptance is calculated as the fraction of events remaining in the combined signal and control
regions after the full event selection.
\label{tab:crosssection_acc}}
\begin{tabular}{lSSSSS}
\toprule
\multirow{2}{*}{Process} &
{\multirow{2}{*}{$\sigma$ $\times$ $\mathcal{B}$ [fb]}} &
\multicolumn{3}{c}{Acceptance [\%]}\\
\cmidrule(lr){3-5}
&  & {0-lepton} & {1-lepton} & {2-lepton}\\
\midrule
$qq\to ZH\to\ell\ell b\bar{b}$  &  29.9  & <0.1     &  0.1 & 6.0 \\
$gg\to ZH\to\ell\ell b\bar{b}$   &   4.8  & <0.1      & 0.2 & 13.5 \\
$qq\to WH\to\ell\nu b\bar{b}$  & 269.0  & 0.2 & 1.0 & {--} \\
$qq\to ZH\to\nu\nu b\bar{b}$ &  89.1  & 1.9 & {--}        & {--} \\
$gg\to ZH\to\nu\nu b\bar{b}$   &   14.3  & 3.5 & {--}        & {--} \\
\bottomrule
\end{tabular}
\end{center}
\end{table}
% End of text imported from the .//paper_04b_acceptance.tex input file
 
\subsection{Multivariate analysis}
\label{subsec:MVA}
Boosted decision trees (BDT) are trained in eight signal regions, corresponding to two jet
categories for the three lepton channels in the high-\ptv\ region, in addition to the two
jet categories for the 2-lepton medium-\ptv\ region. The BDT outputs are used as the
final discriminating variables in the analysis. Two sets of BDTs are constructed with
the same input variables and parameters. The nominal one (BDT$_{VH}$) is designed to separate
Higgs boson events from the sum of expected backgrounds, while the second one (BDT$_{VZ}$) is
used to validate the analysis by the extraction of the diboson $VZ$, $Z\to\bbbar$ process
from the sum of all other SM processes.
 
The same input variables, BDT settings and BDT output binning transformation as those detailed in Ref.~\cite{HIGG-2016-29}
are used, with one exception in the 2-lepton channel where the \met\ is replaced with
$\met/\sqrt{S_{\mathrm{T}}}$ (where
$S_{\mathrm{T}}$ is the scalar sum of transverse momenta of the charged leptons and jets in the
event). Eight to thirteen input variables describing the kinematics of the events are
used depending on the channels, of which \mbb, \ptv\ and $\Delta R(\vec{b_1}, \vec{b_2})$ (where $b_1$ and $b_2$ refer to the two b-tagged jets) are
the most discriminating.
 
\subsection{Dijet-mass analysis}
\label{subsec:dijetmass_selection}
 
A cross-check of the main multivariate analysis is performed by using the invariant mass of
the two $b$-tagged jets as the discriminating variable. Additional selections displayed in
Table~\ref{tbl:evselTableCBA} increase the purity of the signal regions and are necessary
to improve the sensitivity of this method.
 
The high-\ptv\ region is split into two regions $150\,\GeV<\ptv<200\,\GeV$ and
$\ptv>200\,\GeV$, with further requirements placed upon $\Delta R(\vec{b_1}, \vec{b_2})$.
Selections on the transverse mass of the $W$ boson (\mtw) and on $\met/\sqrt{S_{\mathrm{T}}}$
reduce the $\ttbar$ background in the 1- and 2-lepton channels, respectively.
 
In the 1-lepton channel the \mbb\ distribution is able to sufficiently constrain the
$W$+HF background, thus it is not necessary to separate events into a dedicated $W$+HF CR.
 
% The next lines are included from the .//tables/table_04_evtselcba.tex input file
\begin{table}[htbp]
\centering
\caption{Summary of the event selection criteria in the 0-, 1- and 2-lepton channels for the dijet-mass analysis,
applied in addition to those described in Table~\ref{tbl:evselTable} for the multivariate analysis.
\label{tbl:evselTableCBA}}
\begin{tabular}{l  c  c  c}
\toprule
\multicolumn{4}{c}{Channel}  \\
\midrule
Selection & 0-lepton & 1-lepton & 2-lepton  \\
\midrule
\mtw\ & - & $<$ 120~\textrm{\GeV} & -   \\
$\met/\sqrt{S_{\mathrm{T}}}$ & - & - & $<3.5\sqrt{\mathrm{\GeV}}$  \\
\bottomrule
\multicolumn{4}{c}{}\\
\toprule
\multicolumn{4}{c}{$\pt^V$ regions}  \\
\midrule
$\pt^V$ & $75-150$~\textrm{\GeV} & $150-200$~\textrm{\GeV} & $>$ 200~\GeV \\
& (2-lepton only)&  & \\
\midrule
$\Delta R(\vec{b}_1,\vec{b}_2)$ & $<$3.0 & $<$1.8 &$<$1.2  \\
\bottomrule
\end{tabular}
\end{table}
% End of text imported from the .//tables/table_04_evtselcba.tex input file
 
% End of text imported from the .//paper_04_selection.tex input file
 
\section{Data, simulated samples and multi-jet background}
% The next lines are included from the .//paper_03_samples.tex input file
The data used in this analysis were collected at a centre-of-mass energy of 13~\TeV\
during the 2015--2017 running periods.
Events are selected for analysis only if they are of good quality and if all the relevant
detector components are known to have been in good operating condition, which corresponds to a total integrated luminosity of
$79.8\pm 1.6$~\fb~\cite{DAPR-2013-01,Avoni:2018iuv}.
The recorded events contain an average of 32 inelastic $pp$ collisions.
 
Monte Carlo (MC) simulated events are used to model the backgrounds from SM processes and $VH$, $H
\rightarrow b\bar{b}$ signal processes. All simulated processes are normalised using the
most accurate theoretical cross-section predictions currently available and were generated
at least to next-to-leading-order (NLO) accuracy. All samples of simulated events were passed through the ATLAS detector simulation~\cite{SOFT-2010-01} based on \textsc{GEANT~4}~\cite{geant}
and were 	reconstructed with the standard ATLAS reconstruction software. The effects
of multiple interactions in the same and nearby bunch crossings (pile-up) were modelled by overlaying minimum-bias events, simulated using the soft QCD processes of \textsc{Pythia 8.186}~\cite{Pythia8} with the A2~\cite{ATL-PHYS-PUB-2012-003} set of tuned parameters (tune) and \textsc{MSTW2008LO}~\cite{Martin:2009iq} parton distribution functions (PDF). For all samples of simulated events, except for those generated using \textsc{Sherpa}~\cite{sherpa}, the \textsc{EvtGen v1.2.0} program~\cite{Lange:2001uf} was used to describe the decays of bottom and charm hadrons.
A summary of all the generators used for the simulation of the signal and background
processes is shown in Table~\ref{tab:samples}. Samples produced with alternative generators are used to estimate systematic uncertainties in the event modelling, as described in Section~\ref{sec:syst}.
 
% The next lines are included from the .//tables/table_03_samples.tex input file
\begin{table}[tb!]
\begin{center}{\fontsize{7}{7.9}\selectfont
\caption
{The generators used for the simulation of the signal and background
processes. If not specified, the order of the cross-section calculation refers to the expansion in the strong coupling constant ($\alphas$).
The acronyms ME, PS and UE stand for matrix element, parton shower and underlying event,  respectively.
$(\star)$ The events were generated using the first PDF in the NNPDF3.0NLO set and
subsequently reweighted to the PDF4LHC15NLO set~\cite{Butterworth:2015oua} using the internal algorithm in \textsc{Powheg-Box v2}.
$(\dagger)$ The NNLO(QCD)+NLO(EW) cross-section calculation for the $pp \to ZH$ process already includes the $gg\to ZH$ contribution.
The $qq\to ZH$  process is normalised using the cross-section for the $pp \to ZH$ process,
after subtracting the $gg\to ZH$ contribution. An additional scale factor is applied to
the $qq \rightarrow VH$ processes as a function of the transverse momentum of the vector
boson, to account for electroweak (EW) corrections at NLO\@. This makes use of the $VH$ differential cross-section computed with \textsc{Hawk}~\cite{Denner:2011id,Denner:2014cla}.
\protect\label{tab:samples}}
\begin{tabular}{llllll}
\toprule
\hsp Process & ME generator & ME PDF &PS and & UE model & Cross-section \hspace{2.5cm}\\
& & & Hadronisation & tune & order\\
\midrule
\multicolumn{6}{l}{Signal, mass set to 125~\GeV\ and $b\bar{b}$ branching fraction to 58\%} \\
\midrule
\hsp $qq\to WH$  &\textsc{Powheg-Box v2}~\cite{Alioli:2010xd} +&
NNPDF3.0NLO$^{(\star)}$~\cite{Ball:2014uwa} &\PYTHIAV{8.212}~\cite{Pythia8} & AZNLO~\cite{AZNLO:2014} & NNLO(QCD)+  \\
\hspace*{0.5cm} $\to\ell\nu b\bar{b}$&   \textsc{GoSam}~\cite{arXiv:1111.2034} + \textsc{MiNLO}~\cite{arXiv:1206.3572,Luisoni:2013kna}  & & & &NLO(EW)~\cite{Ciccolini:2003jy,Brein:2003wg,Ferrera:2011bk,Brein:2011vx,Ferrera:2013yga,Ferrera:2014lca,Campbell:2016jau} \\
\hsp $qq\to ZH$ &\textsc{Powheg-Box v2} + & NNPDF3.0NLO$^{(\star)}$ &\PYTHIAV{8.212} & AZNLO & NNLO(QCD)$^{(\dagger)}$+  \\
\hspace*{0.5cm}$\to\nu\nu b\bar{b}/\ell\ell b\bar{b}$&   \textsc{GoSam} + \textsc{MiNLO}  & & & &NLO(EW) \\
\hsp $gg  \to ZH$ &\textsc{Powheg-Box v2} & NNPDF3.0NLO$^{(\star)}$ &\PYTHIAV{8.212} & AZNLO & NLO+  \\
\hspace*{0.5cm}$\to\nu\nu b\bar{b}/\ell\ell b\bar{b}$&  & & & &NLL~\cite{Altenkamp:2012sx,Hespel:2015zea,ggzhnll,Harlander:2013mla,Brein:2012ne}\\
\midrule
\multicolumn{6}{l}{Top quark, mass set to 172.5~\GeV}  \\
\midrule
\hsp $t\bar{t}$ &\textsc{Powheg-Box v2}~\cite{Frixione:2007nw} &  NNPDF3.0NLO &\PYTHIAV{8.230} &A14~\cite{ATL-PHYS-PUB-2014-021}& NNLO+NNLL~\cite{Czakon:2011xx} \\
\hsp $s$-channel &\textsc{Powheg-Box v2}~\cite{Alioli:2009je} & NNPDF3.0NLO &\PYTHIAV{8.230} &A14 & NLO~\cite{PhysRevD.81.054028} \\
\hsp $t$-channel &\textsc{Powheg-Box v2}~\cite{Alioli:2009je} & NNPDF3.0NLO &\PYTHIAV{8.230} &A14& NLO~\cite{Kidonakis:2011wy} \\
\hsp $Wt$ &\textsc{Powheg-Box v2}~\cite{Re:2010bp} & NNPDF3.0NLO &\PYTHIAV{8.230} &A14& Approximate NNLO~\cite{PhysRevD.82.054018} \\
\midrule
\multicolumn{6}{l}{Vector boson + jets} \\
\midrule
\hsp $W\to\ell\nu$ & \textsc{Sherpa 2.2.1}~\cite{sherpa,Cascioli:2011va,Gleisberg:2008fv} & NNPDF3.0NNLO &  \textsc{Sherpa 2.2.1}~\cite{Schumann:2007mg,Hoeche:2012yf} & Default & NNLO~\cite{Catani:2009sm}  \\
\hsp $Z/\gamma^{*}\to\ell\ell$ & \textsc{Sherpa 2.2.1} & NNPDF3.0NNLO &  \textsc{Sherpa 2.2.1} & Default & NNLO    \\
\hsp $Z\to\nu\nu$ & \textsc{Sherpa 2.2.1} & NNPDF3.0NNLO &  \textsc{Sherpa 2.2.1} & Default & NNLO    \\
\midrule
\multicolumn{6}{l}{Diboson} \\
\midrule
\hsp $qq \to WW$ & \textsc{Sherpa 2.2.1} & NNPDF3.0NNLO &  \textsc{Sherpa 2.2.1} & Default & NLO  \\
\hsp $qq \to WZ$ & \textsc{Sherpa 2.2.1} & NNPDF3.0NNLO &  \textsc{Sherpa 2.2.1} & Default & NLO  \\
\hsp $qq \to ZZ$ & \textsc{Sherpa 2.2.1} & NNPDF3.0NNLO &  \textsc{Sherpa 2.2.1} & Default & NLO  \\
\hsp $gg \to VV$ & \textsc{Sherpa 2.2.2} & NNPDF3.0NNLO &  \textsc{Sherpa 2.2.2} & Default & NLO  \\
\bottomrule
\end{tabular}}
\end{center}
\end{table}
% End of text imported from the .//tables/table_03_samples.tex input file
 
The background processes involving $W$ or $Z$ boson decays into leptons (including those
in which the $W$ boson arises from a top-quark decay) are collectively referred to in the
following as electroweak (EW) backgrounds and were simulated as described above.
In contrast, the multi-jet background is estimated in all three
channels using data-driven methods. In both the 0- and 2-lepton channels, the multi-jet
contribution is estimated from template fits to data, using the simulated samples to model
the EW backgrounds and a functional form to model the multi-jet background. The template
fit is performed using a variable that provides significant discrimination between the
multi-jet and EW processes, with any selection on that variable removed. In the 0-lepton
channel, $\min [ \Delta\phi(\vecmet, \mathrm{\mathbf{jets}}) ]$ is used, and in the
2-lepton channel, the dilepton mass distribution is used for the case where the charges of the
lepton candidates have the same sign, assuming the multi-jet contribution is
symmetric for opposite- and same-sign lepton charges. In both cases, it is found that the
multi-jet contribution is sufficiently small that it can be neglected in the global
likelihood fit without having any impact on the extracted signal.
 
The multi-jet background is found to be non-negligible in the 1-lepton channel and is
estimated separately in the electron and muon sub-channels, and in the 2- and 3-jet
categories. In each category, a template fit to the transverse mass distribution of the $W$ boson
candidate is performed, which offers the clearest discrimination between the multi-jet and
EW processes, to extract the multi-jet yield. The template used for the multi-jet
contribution is obtained from data in a control region after subtraction of the residual
EW contribution, based on MC predictions, while the template for the EW contribution in
the signal region is obtained directly from MC predictions. The control region is enriched
in multi-jet events that are kinematically close to  the corresponding signal region but
not overlapping with it, and is defined by applying the nominal selection but inverting
the stricter lepton isolation requirements. To increase the statistical precision of the data-driven
estimate, the number of required $b$-tagged jets is reduced from two to one in the
multi-jet enriched control region. The template fit applied in the signal region
determines the normalisation of the multi-jet contribution, while the shape of the BDT
discriminant (or of other relevant observables) is obtained using a control region analogously to the \mtw\ template. Both the normalisation and shape derived for the BDT discriminant are then used in the global likelihood fit.
The multi-jet contribution in the 2-jet category is found to be $1.9\%$ ($2.8\%$) of the
total background contribution in the electron (muon) sub-channel, while in the 3-jet
category it is found to be $0.2\%$ ($0.4\%$). These estimates are subject to sizeable systematic uncertainties, which are described in Section~\ref{sec:syst}.
% End of text imported from the .//paper_03_samples.tex input file

\section{Systematic uncertainties}
\label{sec:syst}
% The next lines are included from the .//paper_06_syst.tex input file
The sources of systematic uncertainty can be broadly divided into four groups: those of experimental nature, those related
to the modelling of the simulated backgrounds, those related to the multi-jet background estimation, and those associated
with the Higgs boson signal simulation. The estimation of the uncertainties closely follows the methodology outlined in Ref.~\autocite{HIGG-2016-29} and is briefly summarised below.
 
\subsection{Experimental uncertainties}
\label{subsec:syst_exp}
 
The dominant experimental uncertainties originate from the $b$-tagging correction factors, determined from the difference between the efficiency measured in data and simulation, from
the jet energy scale corrections and from the modelling of the jet energy resolution.
The $b$-tagging correction factors are derived
separately for $b$-jets, $c$-jets and light-flavour jets~\autocite{PERF-2016-05updated, ATLAS-CONF-2018-001, ATLAS-CONF-2018-006}. All three correction factors have uncertainties estimated from
multiple measurements, which are decomposed into uncorrelated components that are then treated independently, resulting in three uncertainties for $b$-jets and $c$-jets, and five for light-flavour jets.
The approximate size of the uncertainty in the tagging efficiency is 2\% for $b$-jets, 10\% for $c$-jets and 40\% for light-flavour
jets. Additional uncertainties are considered in the extrapolation of the $b$-jet efficiency calibration to jets with $\pt > 300$~\GeV\ and in the misidentification of hadronically decaying $\tau$-leptons as $b$-jets. The uncertainties in the jet energy scale and resolution are based on their respective measurements~\autocite{PERF-2016-04,PERF-2011-04}.
The many sources of uncertainty in the correction of the jet energy scale are decomposed into 23 uncorrelated components that are treated as independent.
An additional specific uncertainty in the energy calibration of $b$- and $c$-jets is considered.
 
Uncertainties in the reconstruction, identification, isolation and trigger efficiencies of muons~\autocite{PERF-2015-10}
and electrons~\autocite{ATLAS-CONF-2016-024}, along with the uncertainty in their energy scale and resolution, are estimated using 13~\TeV\ data. These are found to have only a small impact on the result.
The uncertainties in the energy scale and resolution of the jets and leptons are propagated
to the calculation of \met, which also has additional  uncertainties from the scale, resolution
and reconstruction efficiency of the tracks used to compute the soft term~\autocite{PERF-2016-07}, along with the modelling
of the underlying event. An uncertainty is assigned to the \met\ trigger correction factors, determined from the difference between the trigger efficiency in data and simulation, to account for the
statistical uncertainty in the measured correction factors and for differences between the correction factors determined from $W$~+~jets,
$Z$~+~jets and $t\bar{t}$ events.
The uncertainty in the combined 2015--2017 integrated luminosity is 2.0\%. It is derived,
following a methodology similar to that detailed in Ref.~\cite{DAPR-2013-01}, and using
the LUCID-2 detector for the baseline luminosity measurements \cite{Avoni:2018iuv}.
The average number of interactions per bunch crossing is rescaled by 1.03 to improve agreement between simulation and data, based on the measurement of the visible cross-section in minimum-bias events~\autocite{STDM-2015-05}, and an uncertainty, as large as the correction, is included.
 
\subsection{Simulated sample uncertainties}
 
Modelling uncertainties are derived for the simulated samples and broadly cover three areas: normalisations, acceptance
differences that affect the relative normalisations between analysis regions with a common normalisation, and the
shapes of the differential distributions of the most important kinematic variables. The overall normalisations and associated
uncertainties for the background processes are taken from the currently most accurate calculations as detailed in
Table~\ref{tab:samples}, apart from the main backgrounds whose normalisations are left unconstrained (floated) in the global
likelihood fit. The additional systematic uncertainties in the acceptance differences and in the shapes are derived either
from particle-level comparisons between nominal and alternative simulated samples, or from comparisons with data in control
regions. The particle-level comparisons are cross-checked with detector-level simulations whenever these are available, and
good agreement is found.
The alternative samples were either produced by other generators or by altering the nominal values of generator parameters.
When acceptance uncertainties are estimated, the nominal and alternative samples are normalised
using the same production cross-section. Shape uncertainties are considered in each of the analysis regions separately,
with the samples scaled to have the same normalisation in each region. In this case, the uncertainty is taken from the
alternative sample that differs most in shape from the nominal sample. Shape uncertainties are only
derived for the \mbb\ and \ptv\ variables, as it was found sufficient to only consider the changes induced in
these variables to cover the overall shape variation of the BDT$_{VH}$ discriminant.
Full details are provided in Ref.~\cite{HIGG-2016-29}.
 
\subsubsection{Background uncertainties}
 
The systematic uncertainties affecting the modelling of the background samples are summarised in
Tables~\ref{tab:bkg_systematics} and~\ref{tab:bkgdib_systematics} and key details of the treatment of the backgrounds are
reported below.
 
\begin{table}[p!]
\begin{small}
\begin{center}
\caption
{Summary of the systematic uncertainties in the background modelling for
$Z$~+~jets, $W$~+~jets, \ttb, single top-quark and multi-jet production.
An `S' symbol is used when only a shape uncertainty is assessed.
The regions for which the normalisations float independently are listed in brackets. Where the size of an acceptance systematic uncertainty varies between regions, a range is displayed.
\label{tab:bkg_systematics}}
\begin{tabular}{ l  c}
\toprule
\multicolumn{2}{c}{$Z$~+~jets}\\
\midrule
$Z+ll$ normalisation & 18\% \\
$Z+cl$ normalisation & 23\% \\
$Z$~+~HF normalisation & Floating (2-jet, 3-jet) \\
$Z+bc$-to-$Z+bb$ ratio & 30 -- 40\% \\
$Z+cc$-to-$Z+bb$ ratio & 13 -- 15\% \\
$Z+bl$-to-$Z+bb$ ratio & 20 -- 25\% \\
0-to-2 lepton ratio & 7\% \\
\mbb, \ptv  & S \\
\midrule
\multicolumn{2}{c}{$W$~+~jets}\\
\midrule
$W+ll$ normalisation & 32\% \\
$W+cl$ normalisation & 37\% \\
$W$~+~HF normalisation & Floating (2-jet, 3-jet)\\
$W+bl$-to-$W+bb$ ratio & 26\% (0-lepton) and 23\% (1-lepton) \\
$W+bc$-to-$W+bb$ ratio & 15\% (0-lepton) and 30\% (1-lepton)    \\
$W+cc$-to-$W+bb$ ratio & 10\% (0-lepton) and 30\% (1-lepton) \\
0-to-1 lepton ratio & 5\%  \\
$W$~+~HF CR to SR ratio & 10\% (1-lepton) \\
\mbb, \ptv  & S \\
\midrule
\multicolumn{2}{c}{\ttb\ (all are uncorrelated between the 0+1- and 2-lepton channels)}\\
\midrule
\ttb\ normalisation & Floating (0+1-lepton, 2-lepton 2-jet, 2-lepton 3-jet) \\
0-to-1 lepton ratio & 8\% \\
2-to-3-jet ratio & 9\% (0+1-lepton only) \\
$W$~+~HF CR to SR ratio & 25\% \\
\mbb, \ptv & S \\
\midrule
\multicolumn{2}{c}{Single top-quark}\\
\midrule
Cross-section & 4.6\% ($s$-channel), 4.4\% ($t$-channel), 6.2\% ($Wt$) \\
Acceptance 2-jet & 17\% ($t$-channel), 55\% ($Wt(bb)$), 24\% ($Wt(\mathrm{other})$) \\
Acceptance 3-jet & 20\% ($t$-channel), 51\% ($Wt(bb)$), 21\% ($Wt(\mathrm{other})$)  \\
\mbb, \ptv & S ($t$-channel, $Wt(bb)$, $Wt(\mathrm{other})$) \\
\midrule
\multicolumn{2}{c}{Multi-jet (1-lepton)}\\
\midrule
Normalisation & 60 -- 100\% (2-jet), 90 -- 140\% (3-jet) \\
BDT template & S \\
\bottomrule
\end{tabular}
\end{center}
\end{small}
\end{table}

\begin{table}[ht!]
\begin{small}
\begin{center}
\caption
{Summary of the systematic uncertainties in the background modelling for diboson production. An `S' symbol is used when only a shape uncertainty is assessed and `PS/UE' indicates parton shower/underlying event.
When extracting the $(W/Z)Z$ diboson production signal yield, as the normalisations are unconstrained,
the normalisation uncertainties are removed. Where the size of an acceptance systematic uncertainty varies between regions, a range is displayed.
\label{tab:bkgdib_systematics}}
\begin{tabular}{ l  c}
\toprule
\multicolumn{2}{c}{$ZZ$}\\
\midrule
Normalisation & 20\% \\
0-to-2 lepton ratio & 6\% \\
Acceptance from scale variations& 10 -- 18\% \\
Acceptance from PS/UE variations for 2 or more jets & 6\% \\
Acceptance from PS/UE variations for  3  jets & 7\% (0-lepton), 3\% (2-lepton)\\
\mbb, \ptv,  from  scale variations& S (correlated with $WZ$ uncertainties)\\
\mbb, \ptv, from PS/UE variations& S (correlated with $WZ$ uncertainties)\\
\mbb,  from matrix-element variations& S (correlated with $WZ$ uncertainties)\\
\midrule
\multicolumn{2}{c}{$WZ$}\\
\midrule
Normalisation & 26\% \\
0-to-1 lepton ratio & 11\% \\
Acceptance from scale variations & 13 -- 21\%   \\
Acceptance from PS/UE variations for 2 or more jets & 4\%\\
Acceptance from PS/UE variations for  3  jets & 11\% \\
\mbb, \ptv,  from  scale variations& S (correlated with $ZZ$ uncertainties)\\
\mbb, \ptv, from PS/UE variations& S (correlated with $ZZ$ uncertainties)\\
\mbb,  from matrix-element variations& S (correlated with $ZZ$ uncertainties)\\
\midrule
\multicolumn{2}{c}{$WW$} \\
\midrule
Normalisation & 25\% \\
\bottomrule
\end{tabular}
\end{center}
\end{small}
\end{table}

\textbf{\vjets\ production} The \vjets\ backgrounds are subdivided into three different components based upon the jet flavour labels of the two $b$-tagged jets in the event. The main background contributions ($V+bb$, $V+bc$, $V+bl$ and $V+cc$) are jointly considered as the \vhf\ background. Their overall normalisation, separately in the 2- and 3-jet categories, is free to float in the global likelihood fit. The remaining flavour components, \vcl\ and \vll, constitute less than $\sim1$\% of the background in each analysis region, so only uncertainties in the normalisation of these backgrounds are included. Acceptance uncertainties are estimated for the relative normalisations of the different regions that share a common floating normalisation parameter. In the case of the \whf\ background, this includes the uncertainties in the ratio of the event yield in the 0-lepton channel to that in the 1-lepton channel and, in the 1-lepton channel, in the ratio of the event yield in the \whf\ control region to that in the signal region. For the \zhf\ background, there is an uncertainty in the ratio of the event yield in the 0-lepton channel to that in the 2-lepton channel.
Uncertainties are also estimated in the relative normalisation of the four heavy-flavour
components that constitute the \vhf\ background. These are taken as uncertainties in the
$bc$, $cc$ and $bl$ yields compared with the dominant $bb$ yield and are estimated
separately in each channel in a manner similar to the acceptance systematic uncertainties. Uncertainties are also derived
for the shapes of the \mbb\ and \ptv\ distributions, which are evaluated for \whf\ from comparisons with alternative samples and
for \zhf\ from comparisons with data in \mbb\ sidebands.
 
\textbf{\ttb\ production} Due to the significantly different regions of phase space probed, the \ttb\ background in the 0- and 1-lepton channels
(jointly referred to as 0+1-lepton channel in the following) is considered independently from the \ttb\ background in the 2-lepton
channel; different overall floating normalisation factors are considered, and acceptance uncertainties are derived
separately and taken as uncorrelated between the 0+1- and 2-lepton channels. For the 0+1- lepton channels, uncertainties are
considered in the normalisation ratios of the 2-jet and 3-jet categories, of the $W$~+~HF control and signal regions, and of the 1-lepton and 0-lepton channels.
For the 2-lepton channel, the normalisations in the 2- and 3-jet categories are both left floating, and are effectively determined in their respective $e\mu$ control regions.
Uncertainties in the shapes of the \ptv\ and \mbb\ distributions are estimated in the 0+1- and 2-lepton channels separately
from comparisons with alternative samples. In addition, the modelling of the \ttb\ background is validated in the 2-lepton channel by using the data events from the $e\mu$ control region to model this background in the signal region, with good agreement found.
 
\textbf{Single top-quark production} In the $Wt$- and $t$-channels, uncertainties are derived for the normalisation,
acceptance and shapes of the \mbb\ and \ptv\ distributions. For the $Wt$-channel, the estimated modelling uncertainties are
based on the flavour of the two $b$-tagged jets, due to the different regions of phase space being probed when there are
two $b$-jets ($bb$) present compared with events where there are fewer $b$-jets present (other). Only a normalisation
uncertainty is derived for the $s$-channel, since its contribution is negligible overall.
 
\textbf{Diboson production} The diboson backgrounds are composed of three distinct processes: $WZ$, $WW$ and $ZZ$
production. Given the small contribution from $WW$ production ($<0.1\%$ of the total background) only a normalisation
uncertainty is assigned. The more important contributions from the $WZ$ and $ZZ$ backgrounds have uncertainties derived for
the overall normalisation, the relative acceptance between regions and for the \mbb\ and \ptv\ shapes. These are derived
following the procedure described in Ref.~\autocite{HIGG-2016-29} and are outlined in Table~\ref{tab:bkgdib_systematics}.
 
\subsubsection{Signal uncertainties}
 
The systematic uncertainties that affect the modelling of the signal are summarised in Table~\ref{tab:sig_systematics}.
They are derived following the procedure outlined in Ref.~\autocite{HIGG-2016-29}, but with updated alternative samples
generated with a larger number of events, and using a parameter tune optimized more recently for the evaluation of the parton shower uncertainty. This substantially reduces the parton shower and underlying event (PS/UE) uncertainties. The systematic uncertainties in the calculations of the $VH$ production cross-sections and the
$H\to\bbbar$ branching fraction\footnote{Such systematic uncertainties fully degenerate with the signal yield do not affect the
calculation of the significance relative to the background-only prediction.} are assigned following the recommendations of the LHC Higgs Cross Section Working Group~\cite{Dittmaier:2011ti,Dittmaier:2012vm,Heinemeyer:2013tqa,Harlander:2013mla,Brein:2012ne}.
 
\begin{table}[ht!]
\begin{small}
\begin{center}
\caption
{Summary of the systematic uncertainties in the signal modelling.
An `S' symbol is used when only a shape uncertainty is assessed and `PS/UE' indicates parton shower~/ underlying event. Where the size of an acceptance systematic uncertainty varies between regions, a range is displayed.
\label{tab:sig_systematics}}
\begin{tabular}{ l  c}
\toprule
\multicolumn{2}{c}{Signal}\\
\midrule
Cross-section (scale) & 0.7\% ($qq$), 27\% ($gg$) \\
Cross-section (PDF) & 1.9\% ($qq\rightarrow WH$), 1.6\% ($qq\rightarrow ZH$), 5\% ($gg$) \\
$H \to \bbbar$ branching fraction & 1.7\% \\
Acceptance from scale variations & 2.5 -- 8.8\%   \\
Acceptance from PS/UE variations for 2 or more jets & 2.9 -- 6.2\% (depending on lepton channel)\\
Acceptance from PS/UE variations for  3  jets & 1.8 -- 11\% \\
Acceptance from PDF+$\alphas$ variations & 0.5 -- 1.3\% \\
\mbb, \ptv, from scale variations& S \\
\mbb, \ptv, from PS/UE variations& S \\
\mbb, \ptv, from PDF+$\alphas$ variations & S \\
\ptv\ from NLO EW correction & S \\
\bottomrule
\end{tabular}
\end{center}
\end{small}
\end{table}
 
\subsection{Multi-jet background uncertainties}
 
Systematic uncertainties can have an impact on the data-driven multi-jet estimate used in the 1-lepton channel in two
ways: either changing the \mtw\ distributions used in the multi-jet template fits,
thus impacting the extracted multi-jet normalisations, or directly changing the
multi-jet BDT distributions used in the global likelihood fit. Several
uncertainties are considered, uncorrelated between the electron and muon sub-channels. The
respective variations are added in quadrature for the normalisations, or considered as
separate shape uncertainties. Variations are obtained by changing the definition of
the multi-jet control region (more stringent isolation requirements, a
different single-electron trigger to probe a potential trigger bias in the isolation
requirements), and varying
the normalisation of the contamination from the top ($t\bar{t}$ and $Wt$) and \vjets\
processes in the multi-jet control region. In addition, the following systematic
uncertainties have an impact only on the multi-jet normalisation: use
of another discriminant variable instead of \mtw\ for the template fit (the azimuthal separation between the directions of the lepton transverse momentum and the vectorial sum of the momenta of the two or three jets) and, for the
electron sub-channel only, the inclusion of the $\met<30$~\GeV\ region, which
significantly enhances the multi-jet contribution in the template fit.

% End of text imported from the .//paper_06_syst.tex input file

\section{Statistical analysis}
% The next lines are included from the .//paper_07_stat.tex input file
 
The statistical procedure is based on a likelihood function $\mathcal{L}(\mu,
\vec{\theta})$, constructed as the product of Poisson probability terms over the bins
of the input distributions. The parameter of interest, $\mu$, is the signal strength that
multiplies the SM Higgs boson production cross-section times the branching fraction into
$\bbbar$ and is extracted by maximising the likelihood. Systematic uncertainties enter the likelihood as nuisance parameters (NP),
$\vec{\theta}$. Most of the uncertainties discussed in
Section~\ref{sec:syst} are constrained with Gaussian or log-normal probability density
functions. The normalisations of the largest backgrounds, $t\bar{t}$, $W$+HF and $Z$+HF, can be reliably determined by the
fit, so they are left unconstrained in the likelihood. The uncertainties due to the
limited number of events in the simulated samples used for the background predictions are
included using the Beeston--Barlow technique~\autocite{Barlow:1993dm}. As detailed in
Ref.~\cite{HIGG-2013-23}, systematic variations that are subject to large statistical
fluctuations are smoothed, and systematic uncertainties that have a negligible impact on
the final results are pruned away region-by-region.
 
The probability that the background-only hypothesis is compatible with the observed
data is determined using the $q_0$ test statistic constructed from the profile-likelihood
ratio with the asymptotic approximation~\autocite{Cowan:2010js}.
 
\subsection{Multivariate analysis}
 
As discussed in Section~\ref{subsec:MVA}, the global likelihood fit comprises eight signal
regions, defined as the 2- and 3-jet categories in the high-\ptv\ region for the three
channels, and in the medium-\ptv\ region for the 2-lepton channel. The BDT$_{VH}$
multivariate discriminant output distributions in these regions are input to the fit. The event yields are used in the two
$W$+HF control regions of the 1-lepton channel. In the four $e\mu$ control regions of the
2-lepton channel, the \mbb\ distributions are input to the fit, except for the 2-jet category of the high-\ptv\
region, where the event yield is used. The post-fit normalisation
factors of the unconstrained backgrounds in the global likelihood fit to the 13\,\TeV\ data are shown in Table~\ref{tab:resultscalefactorsMVA}.
 
\begin{table}[tb!]
\begin{center}
\caption{Factors applied to
the nominal normalisations of the \ttb, \whf\, and \zhf\ backgrounds,
as obtained from the global likelihood fit to the 13~\TeV\ data for the nominal
multivariate analysis, used to extract the Higgs boson signal.
The errors represent the combined statistical and systematic uncertainties.
\label{tab:resultscalefactorsMVA}}
\begin{tabular}{lc}
\toprule
Process & Normalisation factor \\
\midrule
\ttb\ 0- and 1-lepton  & $0.98 \pm 0.08$ \\
\ttb\ 2-lepton 2-jet  & $1.06 \pm 0.09$ \\
\ttb\ 2-lepton 3-jet  & $0.95 \pm 0.06$ \\
\whf\ 2-jet &  $1.19 \pm 0.12$ \\
\whf\ 3-jet &  $1.05 \pm 0.12$ \\
\zhf\ 2-jet &  $1.37 \pm 0.11$ \\
\zhf\ 3-jet &  $1.09 \pm 0.09$ \\
\bottomrule
\end{tabular}
\end{center}
\end{table}
 
The effects of systematic uncertainties on the measurement of the signal strength are
displayed in Table~\ref{tab:syst_unc}.
The impact of a category of systematic uncertainties is defined as the
difference in quadrature between the uncertainty in $\mu$ computed when
all NPs are fitted and that when the NPs in the category are fixed to
their best-fit values.
The total statistical
uncertainty is defined as the uncertainty in $\mu$ when all the NPs are fixed to their
best-fit values. The total systematic uncertainty is then defined as the difference in
quadrature between the total uncertainty in $\mu$ and the total statistical uncertainty.
As shown in the table, the systematic uncertainties due to the modelling of the signal play a dominant role,
followed by the uncertainty due to the limited size of the simulated samples, the
modelling of the backgrounds and the $b$-tagging uncertainty.
 
\begin{table}[tb]
\caption{Breakdown of the contributions to the uncertainty in $\mu$.
The sum in quadrature of the systematic uncertainties attached to the
categories differs from the total systematic uncertainty due to correlations.
\label{tab:syst_unc}}
\begin{center}
\begin{tabular}{ l | l  c }
\toprule
\multicolumn{2}{l}{Source of uncertainty} & $\sigma_\mu$ \\
\midrule
\multicolumn{2}{l}{Total} & 0.259 \\
\multicolumn{2}{l}{Statistical} & 0.161 \\
\multicolumn{2}{l}{Systematic} & 0.203 \\
\midrule
\multicolumn{3}{l}{Experimental uncertainties}\\
\midrule
\multicolumn{2}{l}{Jets} &  0.035 \\
\multicolumn{2}{l}{\met} & 0.014 \\
\multicolumn{2}{l}{Leptons} & 0.009 \\
\multirow{3}{*}{$b$-tagging~~~} & $b$-jets & 0.061 \\
& $c$-jets &  0.042 \\
& light-flavour jets & 0.009 \\
& extrapolation & 0.008 \\
\multicolumn{2}{l}{Pile-up} & 0.007 \\
\multicolumn{2}{l}{Luminosity} & 0.023 \\
\midrule
\multicolumn{3}{l}{Theoretical and modelling uncertainties}\\
\midrule
\multicolumn{2}{l}{Signal} & 0.094 \\
\multicolumn{3}{l}{} \\
\multicolumn{2}{l}{Floating normalisations} & 0.035 \\
\multicolumn{2}{l}{$Z$~+~jets} & 0.055 \\
\multicolumn{2}{l}{$W$~+~jets} & 0.060 \\
\multicolumn{2}{l}{\ttb} & 0.050 \\
\multicolumn{2}{l}{Single top quark} & 0.028 \\
\multicolumn{2}{l}{Diboson} & 0.054 \\
\multicolumn{2}{l}{Multi-jet} & 0.005 \\
\multicolumn{3}{l}{} \\
\multicolumn{2}{l}{MC statistical} & 0.070 \\
\bottomrule
\end{tabular}
\end{center}
\end{table}

\subsection{Dijet-mass analysis}
 
In the dijet-mass analysis, the number of signal regions is increased to fourteen as a
consequence of splitting the event regions with \ptv$>150$~\GeV\ in two, while the
$W$+HF CRs are merged into the corresponding SR, as outlined in Section~\ref{subsec:dijetmass_selection}. The \mbb\ distributions are input to the fit
in all categories, except for the 2-jet medium- and high-\ptv\ categories of the 2-lepton $e\mu$ control region, where the event yield is used.
 
\subsection{Diboson analysis}  \label{subsec:dibosonfit}
 
In the diboson analysis, a measurement of the signal strength of the $ZZ$ and $WZ$
processes is conducted to validate the main multivariate analysis. The method differs
from the global likelihood fit only by the use of the BDT$_{VZ}$ output distributions as inputs, instead of BDT$_{VH}$. The parameter of
interest, $\mu_{VZ}$, is the signal strength of the combined $WZ$ and $ZZ$ diboson processes, and the SM Higgs boson
is included as a background process normalised to the predicted SM cross-section with an
uncertainty of 50\%, which conservatively encompasses the previous measurement and uncertainty~\cite{HIGG-2016-29}.
 
\subsection{Combinations} \label{subsec:fitcombination}
\subsubsection{Run 1}

The results of the statistical analysis of the 13 \TeV\ data are combined with those from
the data recorded at 7~\TeV\ and 8~\TeV~\autocite{HIGG-2013-23} to improve the precision
of the measurement. Detailed studies of the impact of the correlation of systematic
uncertainties between the two analyses are reported in Ref.~\autocite{HIGG-2016-29}. In
most cases, the impact of correlations was found to be negligible. Only a $b$-jet-specific
jet energy scale, and theory uncertainties in the Higgs boson signal (overall
cross-section, branching fraction and \ptv-dependent NLO EW corrections) are correlated
across the different centre-of-mass energies.
 
\subsubsection{$H\to\bbbar$}
 
A second combination is performed with the results of the searches for the $H\to\bbbar$
decay in the $\ttbar H$~\autocite{HIGG-2013-27,HIGG-2017-03} and
VBF~\autocite{HIGG-2014-12,HIGG-2016-30} production modes carried out with the Run
1 and Run 2 data. As the analysis targeting the VBF production mode has a sizeable
contribution from gluon--gluon fusion events, it is referred to as the VBF+ggF analysis in the
following.
Constraining the cross-sections of the production modes to be as predicted by the SM, the
combination measures the ratio of the branching fraction of the Higgs boson into $b$-quarks to
the SM prediction.
The only NP correlated across the six analyses is the $H \to \bbbar$
branching fraction that affects the SM prediction. A few other NPs are correlated across some of the analyses, following the
studies conducted for the combinations of Run 1 results~\autocite{HIGG-2015-07},
of analyses of the $\ttbar H$ production mode~\autocite{HIGG-2018-13}, and of Run~2
results.
 
\subsubsection{$VH$}
 
A third combination is also performed combining the Run~2 $VH$, $H\to\bbbar$ result with
other results in the $VH$ production mode, but for the case
of the Higgs boson decaying into two photons or via $ZZ^*$ into four leptons.
 
The measurement of $VH$ production in the $H{\rightarrow}\gamma\gamma$ channel, which uses
five reconstruction-level categories to target leptonic decays of the vector boson, and
two categories targeting hadronic decays of the vector boson, as described in
Ref.~\cite{HIGG-2016-21}, is updated using 79.8~\fb\ of data. Photons are reconstructed from calorimeter energy clusters formed using an enhanced dynamical, topological cell-clustering-based algorithm~\cite{ATL-PHYS-PUB-2017-022}. The signal yield is extracted in each category using a fit to the diphoton invariant mass distribution in the range 105--160\,\GeV. Contamination in these categories from non-$VH$ Higgs boson production is constrained using separate categories designed to measure the $t\bar{t} H$~\cite{HIGG-2018-13}, VBF, and ggF production modes.
 
The measurement of $VH$ production in the four-lepton final state, $H \to
ZZ^{*} \to 4\ell$, where $\ell = e\text{ or }\mu$, was performed with 36.1~\fb~\cite{HIGG-2016-22} and has now been extended to 79.8~\fb. The main enhancements are: improved electron reconstruction~\cite{ATL-PHYS-PUB-2017-022} and an additional event category
targeting vector-boson decays that include missing transverse momentum due to the presence of one or two neutrinos in the final state.  This results in three $VH$ categories, targeting the hadronic decays of the vector boson, charged leptonic decays of the vector boson and decays of the vector boson containing one or more neutrinos.
 
The combination is undertaken as outlined in
Ref.~\cite{HIGG-2014-06}. Constraining the branching fractions for the $ZZ^*$, diphoton
and $\bbbar$ decays to be as predicted by the SM, this combination measures the signal strength of the $VH$ production
mode.
 
% End of text imported from the .//paper_07_stat.tex input file

\section{Results}
% The next lines are included from the .//paper_08_results.tex input file
 
\subsection{Results of the SM Higgs boson search at $\sqrt{s}=13$~\TeV}
 
Figure~\ref{fig:mva_output} shows the BDT output distributions in the most sensitive,
high-\ptv, region.
The background prediction in all post-fit distributions is obtained by normalising the backgrounds and setting the nuisance parameters according to the results of the signal extraction fit. The post-fit signal and background yields are shown in Table~\ref{tbl:yieldTable} for all signal regions.
 
% The next lines are included from the .//tables/table_08_yields.tex input file
\begin{sidewaystable}[phtb]
\small
\centering
\sisetup{round-mode=places, round-precision=0, group-digits=integer, group-minimum-digits=6}
\caption{The Higgs boson signal, background and data yields for each signal region category in each channel after the full selection of the multivariate analysis.
The signal and background yields are normalised to the results of the global likelihood fit.
All systematic uncertainties are included in the indicated uncertainties.
An entry of ``--'' indicates that a specific background component is negligible in a certain region, or that no simulated events are left after the analysis selection.\label{tbl:yieldTable}}
\begin{tabular}{l
S[table-alignment=right,table-format=4.0]@{\,$\pm$\,}S[table-alignment=left,table-format=3.0]
S[table-alignment=right,table-format=5.0]@{\,$\pm$\,}S[table-alignment=left,table-format=3.0]
S[table-alignment=right,table-format=5.0]@{\,$\pm$\,}S[table-alignment=left,table-format=3.0]
S[table-alignment=right,table-format=5.0]@{\,$\pm$\,}S[table-alignment=left,table-format=4.0]
S[table-alignment=right,table-format=5.0]@{\,$\pm$\,}S[table-alignment=left,table-format=3.0]
S[table-alignment=right,table-format=5.0]@{\,$\pm$\,}S[table-alignment=left,table-format=3.0]
S[table-alignment=right,table-format=4.0]@{\,$\pm$\,}S[table-alignment=left,table-format=3.0]
S[table-alignment=right,table-format=4.0]@{\,$\pm$\,}S[table-alignment=left,table-format=3.0] }
\toprule
\multirow{2}{*}{  } & \multicolumn{4}{c}{0-lepton} & \multicolumn{4}{c}{1-lepton} & \multicolumn{8}{c}{2-lepton} \\
& \multicolumn{4}{c}{$\ptv>150\,\GeV$, 2-$b$-tag} & \multicolumn{4}{c}{$\ptv>150\,\GeV$, 2-$b$-tag} & \multicolumn{4}{c}{$75\,\GeV<\ptv<150\,\GeV$, 2-$b$-tag} & \multicolumn{4}{c}{$\ptv>150\,\GeV$, 2-$b$-tag}\\
\midrule
Process  & \multicolumn{2}{c}{2-jet} & \multicolumn{2}{c}{3-jet} & \multicolumn{2}{c}{2-jet} & \multicolumn{2}{c}{3-jet} & \multicolumn{2}{c}{2-jet} & \multicolumn{2}{c}{$\geq$3-jet} & \multicolumn{2}{c}{2-jet} & \multicolumn{2}{c}{$\geq$3-jet}\\
\midrule
$Z+ll$ & 17 & 11 & 27 & 18 & 1.5 & 1.0 & 3.4 & 2.3 & 13.7 & 8.7 & 49 & 32 & 4.1 & 2.8 & 30 & 19\\
$Z+cl$ & 45 & 18 & 76 & 30 & 3.0 & 1.2 & 6.9 & 2.8 & 43 & 17 & 170 & 67 & 11.5 & 4.6 & 88 & 35\\
$Z$~+~HF & 4770 & 140 & 5940 & 300 & 179.5 & 9.1 & 348 & 21 & 7400 & 120 & 14160 & 220 & 1421 & 34 & 5370 & 100\\
$W+ll$ & 20 & 13 & 32 & 22 & 31 & 23 & 65 & 48 & \multicolumn{2}{c}{$<1$} & \multicolumn{2}{c}{$<1$} & \multicolumn{2}{c}{$<1$} & \multicolumn{2}{c}{$<1$}\\
$W+cl$ & 43 & 20 & 83 & 38 & 139 & 67 & 250 & 120 & \multicolumn{2}{c}{$<1$} & \multicolumn{2}{c}{$<1$} & \multicolumn{2}{c}{$<1$} & \multicolumn{2}{c}{$<1$}\\
$W$~+~HF & 1000 & 87 & 1990 & 200 & 2660 & 270 & 5400 & 670 & 1.8 & 0.2 & 13.2 & 1.5 & 1.4 & 0.2 & 4.0 & 0.5\\
Single top quark & 368 & 53 & 1410 & 210 & 2080 & 290 & 9400 & 1400 & 188 & 89 & 440 & 200 & 23.1 & 7.3 & 93 & 26\\
$\ttbar$ & 1333 & 82 & 9150 & 400 & 6600 & 320 & 50200 & 1400 & 3170 & 100 & 8880 & 220 & 104 & 6 & 839 & 40\\
Diboson & 254 & 49 & 318 & 90 & 178 & 47 & 330 & 110 & 152 & 32 & 355 & 68 & 52 & 11 & 196 & 35\\
Multi-jet $e$ sub-ch. & \multicolumn{2}{c}{--} & \multicolumn{2}{c}{--} & 100 & 100 & 41 & 35 & \multicolumn{2}{c}{--} & \multicolumn{2}{c}{--} & \multicolumn{2}{c}{--} & \multicolumn{2}{c}{--}\\
Multi-jet $\mu$ sub-ch. & \multicolumn{2}{c}{--} & \multicolumn{2}{c}{--} & 138 & 92 & 260 & 270 & \multicolumn{2}{c}{--} & \multicolumn{2}{c}{--} & \multicolumn{2}{c}{--} & \multicolumn{2}{c}{--}\\
\midrule
Total bkg. & 7850 & 90 & 19020 & 140 & 12110 & 120 & 66230 & 270 & 10960 & 100 & 24070 & 150 & 1620 & 30 & 6620 & 80\\
\midrule
Signal (post-fit) & 128 & 28 & 128 & 29 & 131 & 30 & 125 & 30 & 51 & 11 & 86 & 22 & 27.7 & 6.1 & 67 & 17\\
\midrule
Data &
\multicolumn{1}{c@{\,\hphantom{$\pm$}\,}}{\tablenum[table-alignment=right,table-format=4.0]{8003}} &&
\multicolumn{1}{c@{\,\hphantom{$\pm$}\,}}{\tablenum[table-alignment=right,table-format=5.0]{19143}} &&
\multicolumn{1}{c@{\,\hphantom{$\pm$}\,}}{\tablenum[table-alignment=right,table-format=5.0]{12242}} &&
\multicolumn{1}{c@{\,\hphantom{$\pm$}\,}}{\tablenum[table-alignment=right,table-format=5.0]{66348}} &&
\multicolumn{1}{c@{\,\hphantom{$\pm$}\,}}{\tablenum[table-alignment=right,table-format=5.0]{11014}} &&
\multicolumn{1}{c@{\,\hphantom{$\pm$}\,}}{\tablenum[table-alignment=right,table-format=5.0]{24197}} &&
\multicolumn{1}{c@{\,\hphantom{$\pm$}\,}}{\tablenum[table-alignment=right,table-format=4.0]{1626}} &&
\multicolumn{1}{c@{\,\hphantom{$\pm$}\,}}{\tablenum[table-alignment=right,table-format=4.0]{6686}}
\\
 
\bottomrule
\end{tabular}
\end{sidewaystable}
% End of text imported from the .//tables/table_08_yields.tex input file
 
\begin{figure}[htbp]
\centering
\begin{tabular}{cccc}
\includegraphics[width=0.4\linewidth]{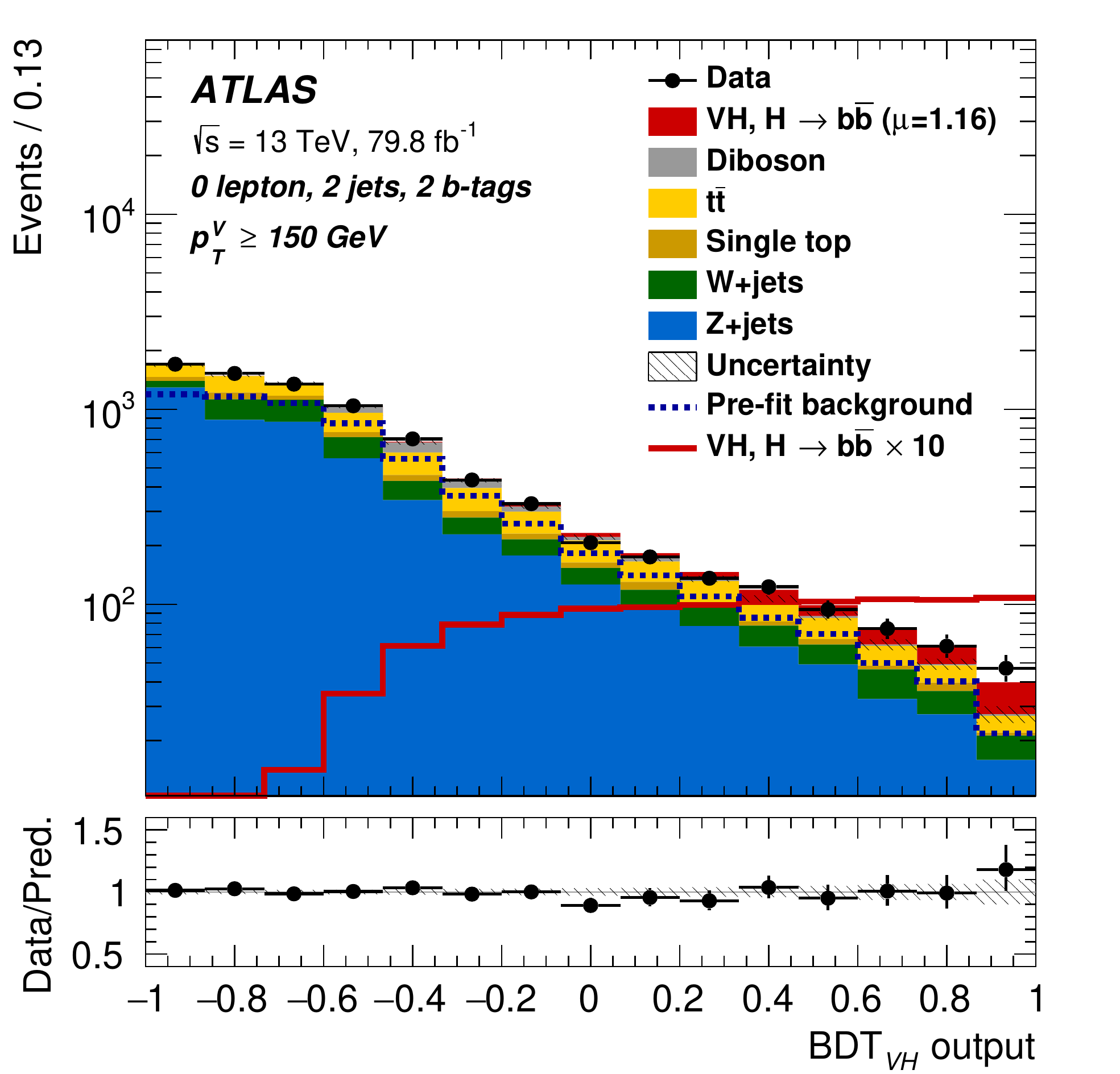}
\includegraphics[width=0.4\linewidth]{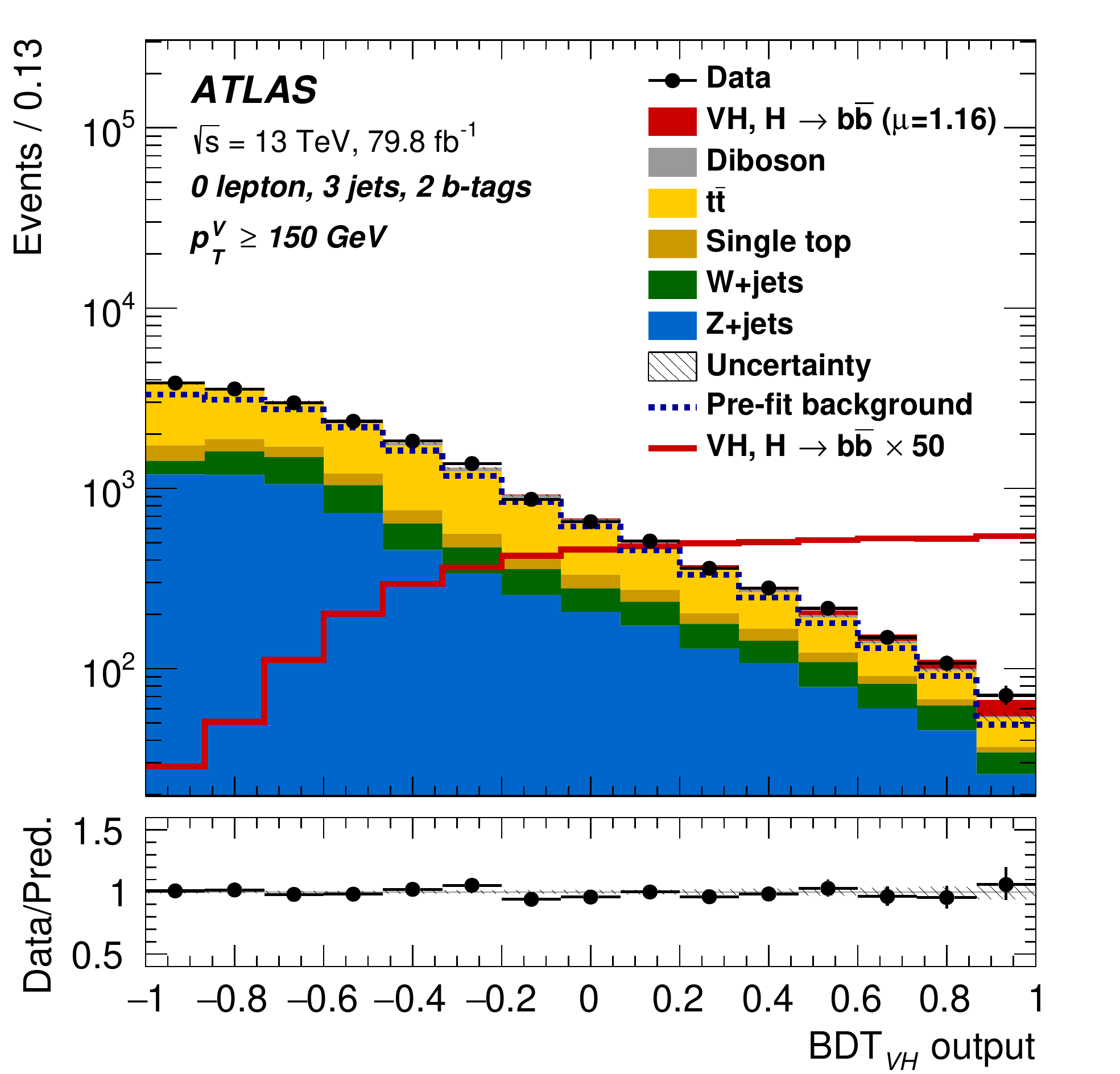}
\\
\includegraphics[width=0.4\linewidth]{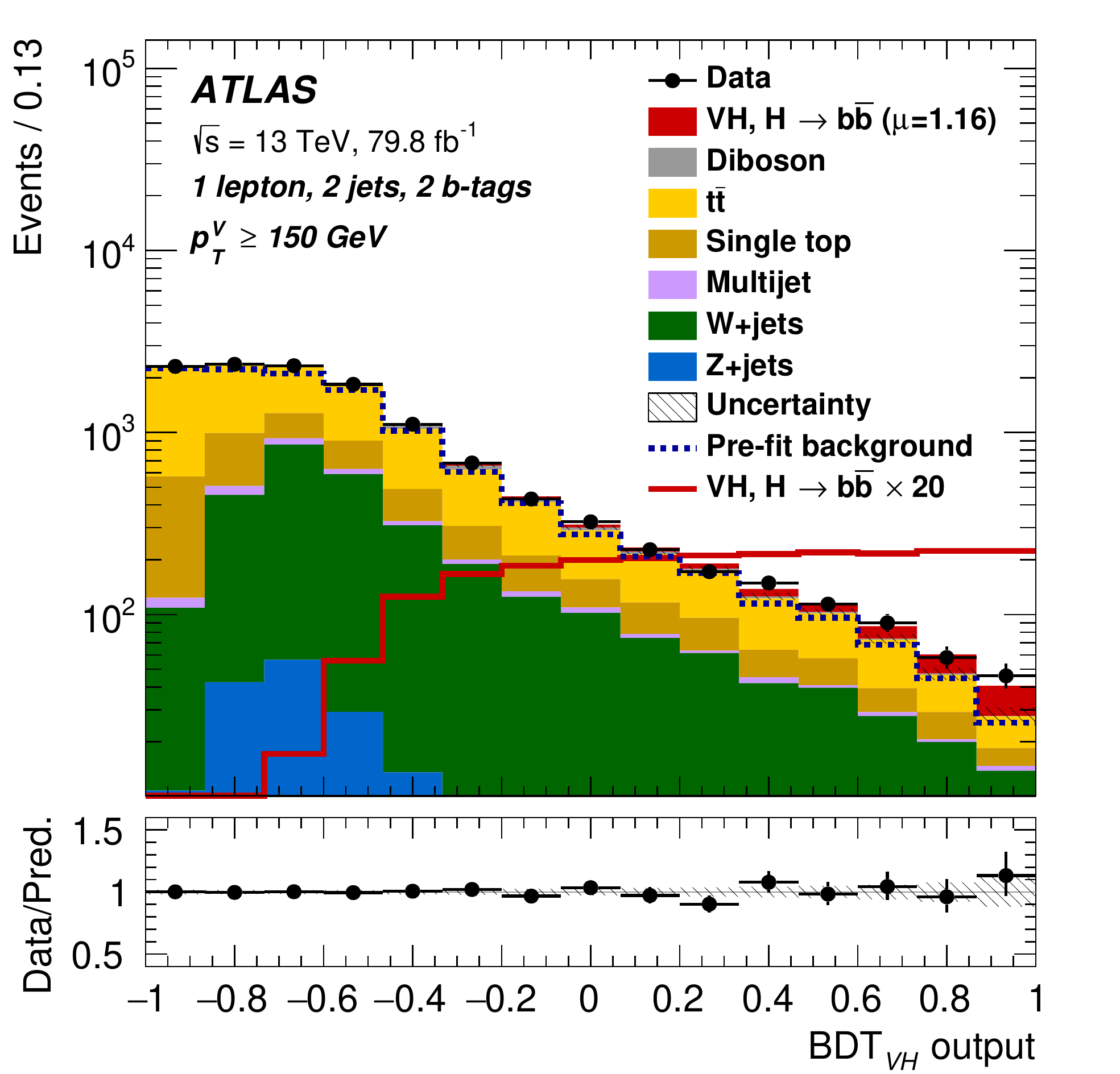}
\includegraphics[width=0.4\linewidth]{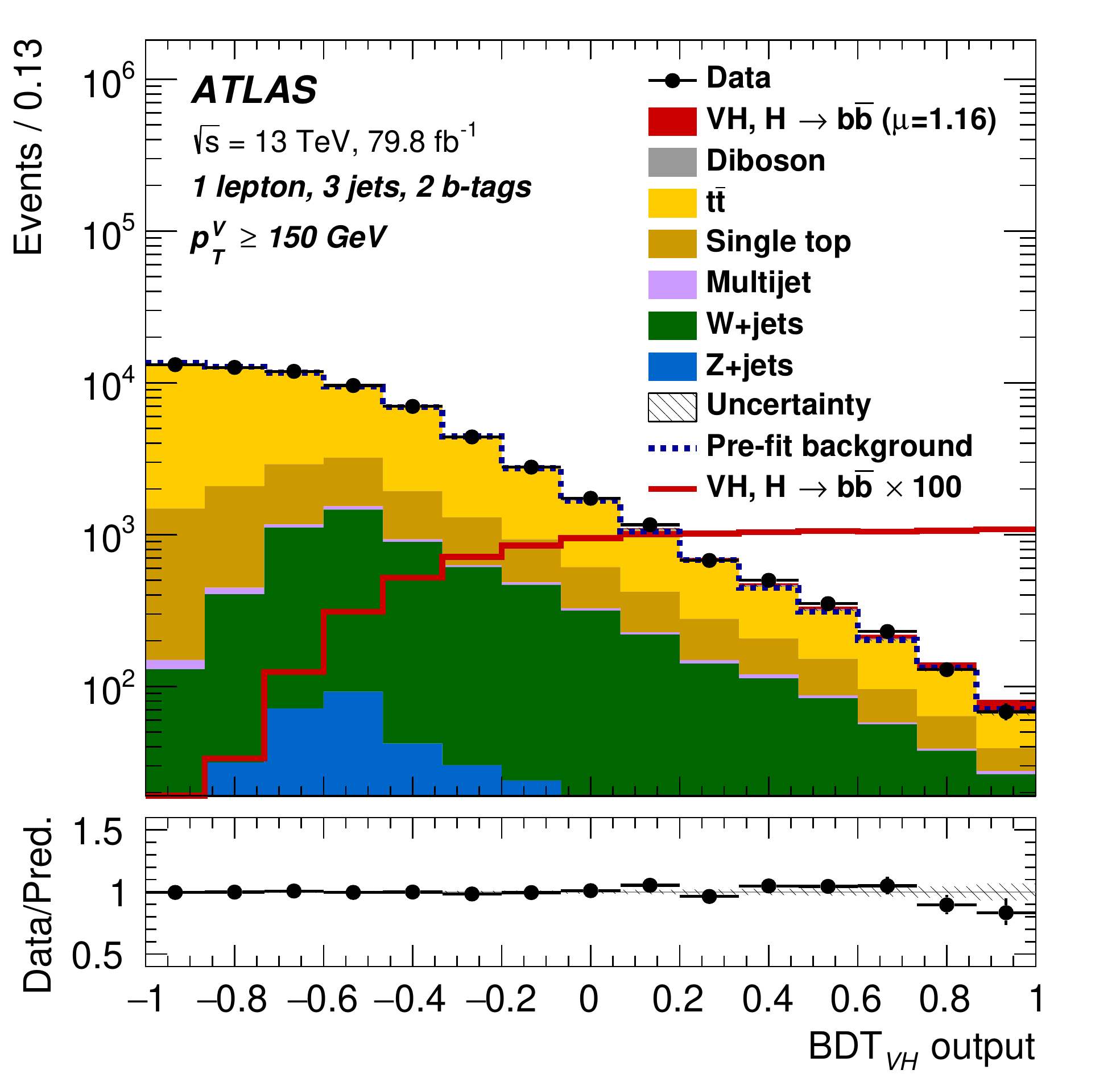}
\\
\includegraphics[width=0.4\linewidth]{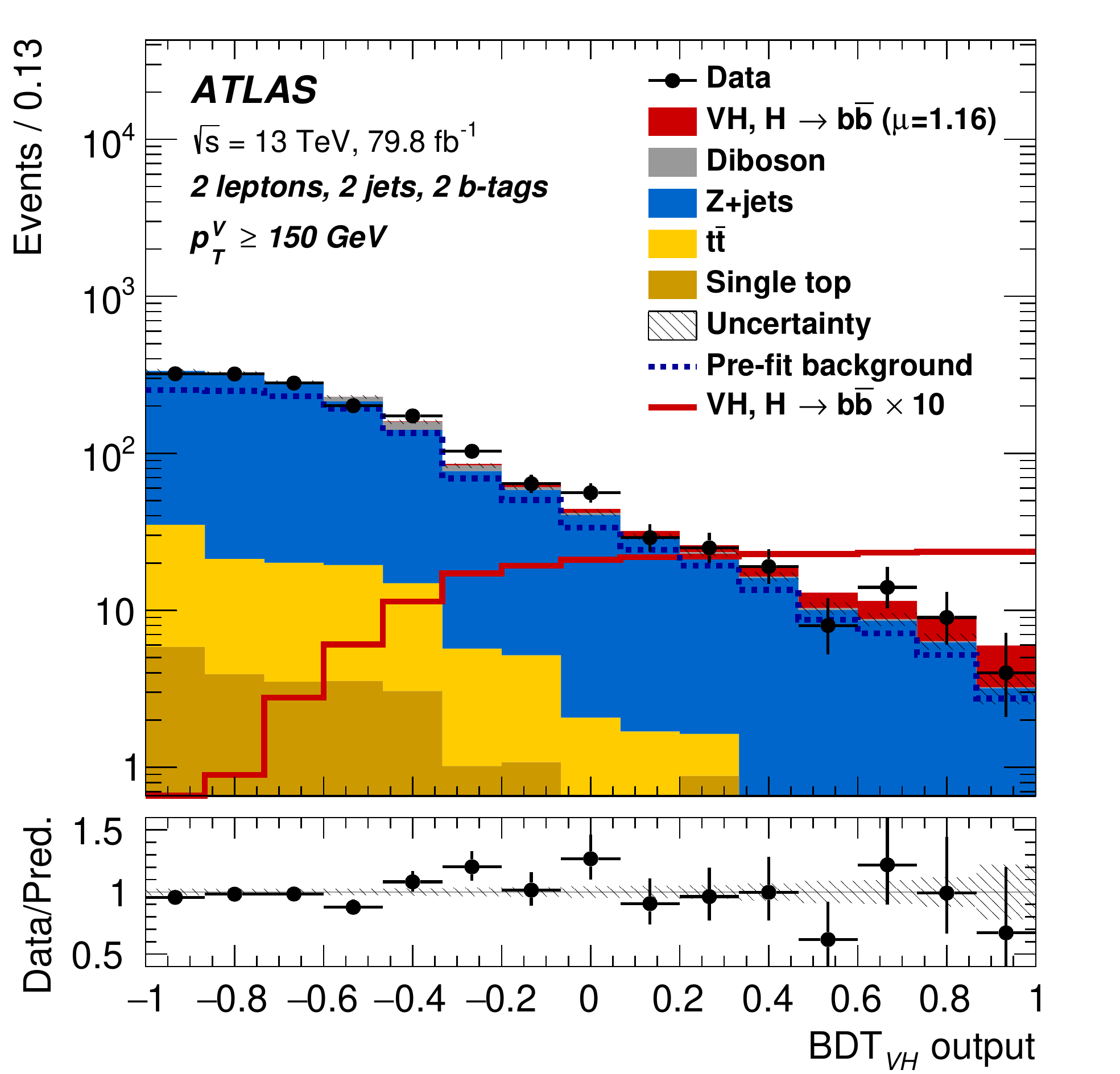}
\includegraphics[width=0.4\linewidth]{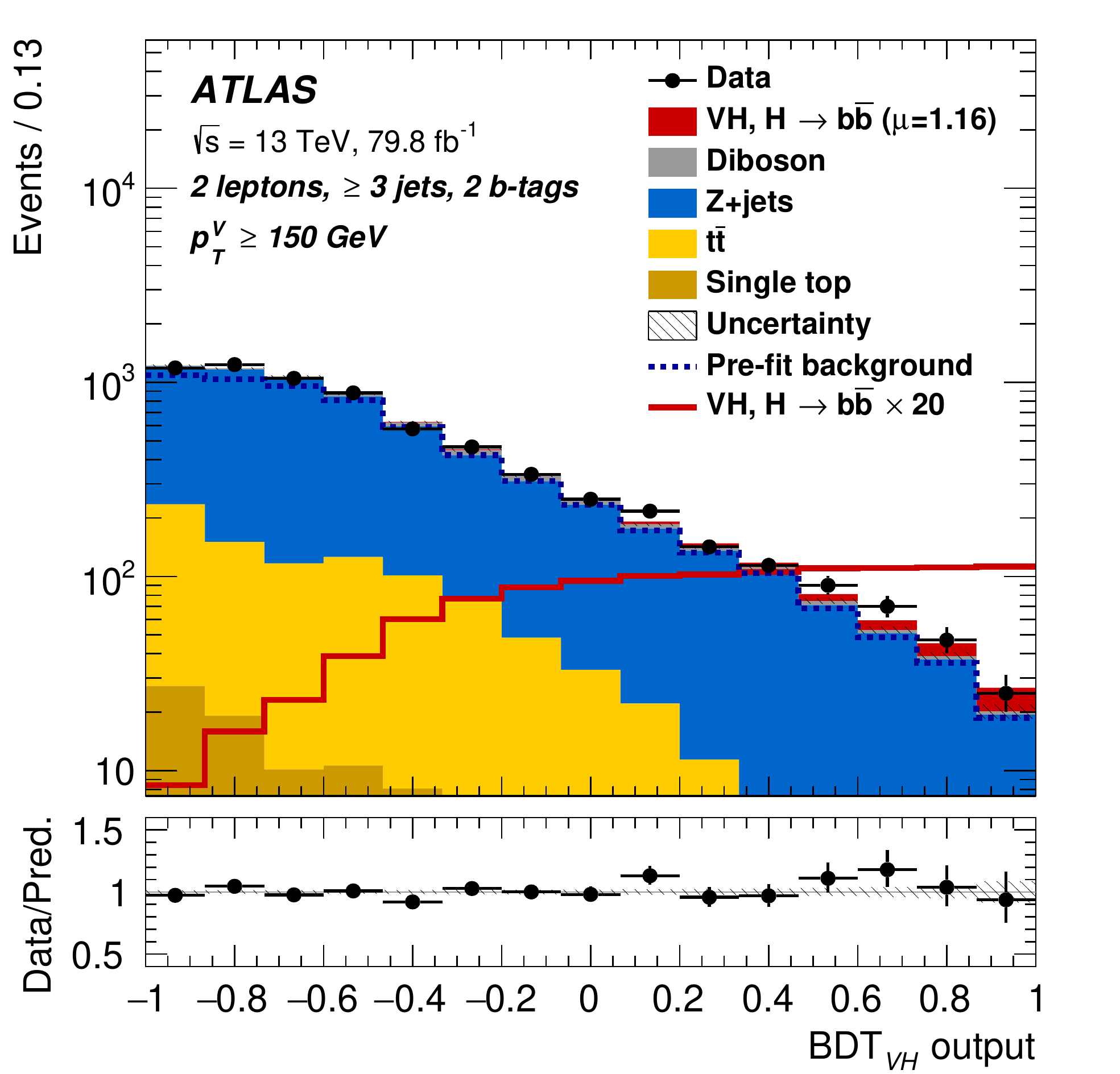}
\\
\end{tabular}
\caption{The BDT$_{VH}$ output post-fit distributions in the 0-lepton~(top), 1-lepton~(middle) and 2-lepton~(bottom) channels for 2-$b$-tag events, in the 2-jet~(left) and exactly 3-jet (or $\ge 3$ jets for the 2-lepton case) (right) categories in the high-\ptv\ region. The background contributions after the global likelihood fit are shown as filled histograms. The Higgs boson signal ($\mh = 125$~\GeV) is shown as a filled histogram on top of
the fitted backgrounds normalised to the signal yield extracted from data ($\mu=1.16$),
and unstacked as an unfilled histogram, scaled by the factor indicated in the legend.
The dashed histogram shows the total pre-fit background. The size of the combined statistical and systematic uncertainty for the
sum of the fitted signal and background is indicated by the hatched band. The ratio
of the data to the sum of the fitted signal ($\mu=1.16$) and background is shown in the lower panel.
The BDT$_{VH}$ output distributions are shown with the binning used in the global likelihood fit. \label{fig:mva_output}}
\end{figure}
 
For a Higgs boson mass of 125~\GeV, when all lepton channels are combined, the
probability $p_0$ of obtaining a signal at least as strong as
the observation from background alone is $5.3\cdot 10^{-7}$, whilst the expected value is $7.3\cdot10^{-6}$. The observation corresponds to an
excess with a significance of 4.9 standard deviations, to be compared with an expectation of
4.3 standard deviations. The fitted value of the signal strength is:
\begin{eqnarray*}
\mu_{VH}^{bb} = 1.16 ^{+0.27}_{-0.25} = 1.16 \pm 0.16 \mathrm{(stat.)} ^{+0.21}_{-0.19} \mathrm{(syst.)}.
\end{eqnarray*}
 
Figure~\ref{fig:logstob} shows the data, background and signal yields, where final-discriminant bins in all regions are combined into bins of $\log_{10}(S/B)$. Here, $S$ and $B$ are the fitted signal and background yields in each analysis bin, respectively.
 
\begin{figure}[thb!]
\begin{center}
\includegraphics[width=0.7\textwidth]{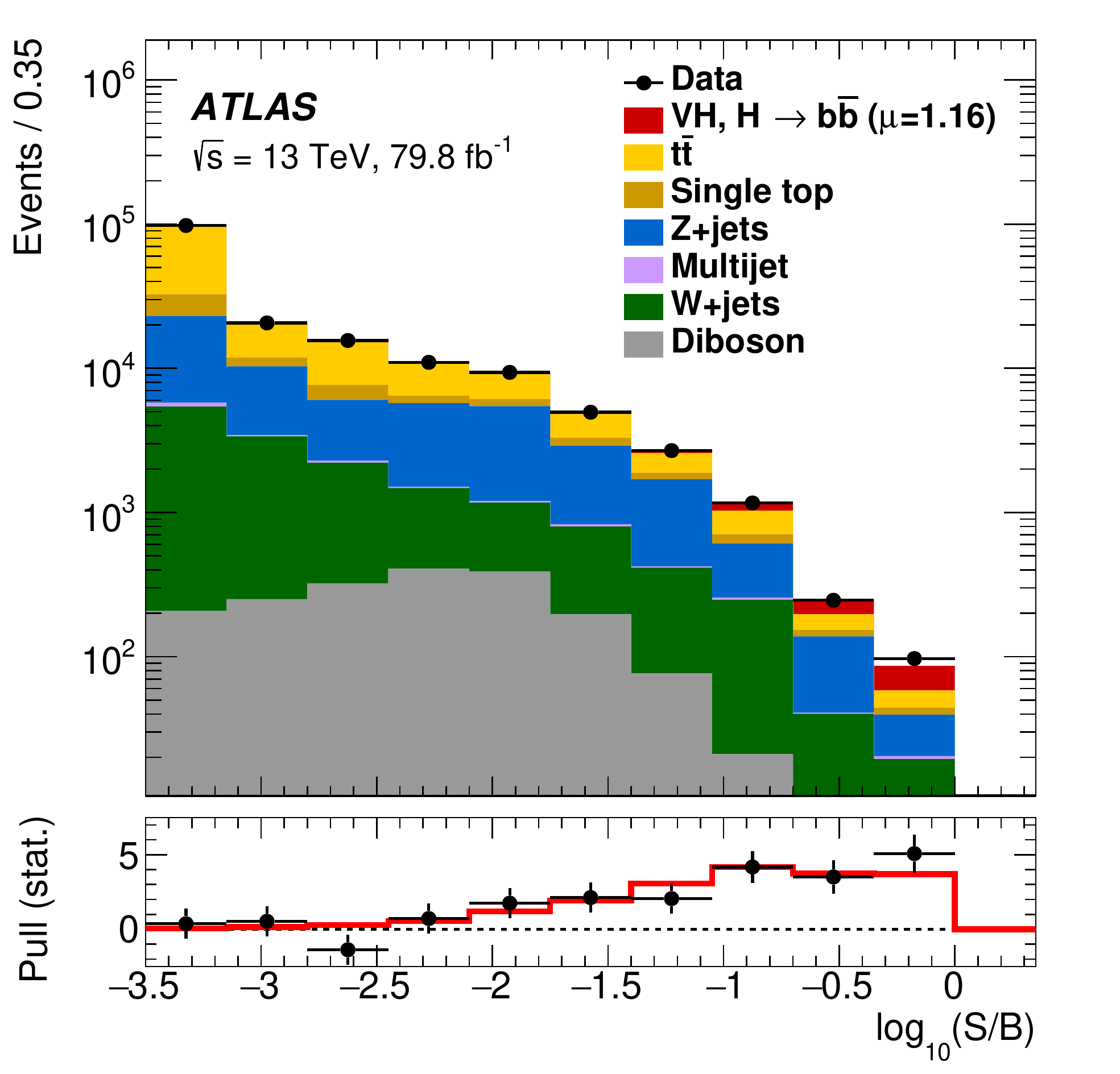}
\end{center}
\caption{Event yields as a function of $\log_{10}(S/B)$ for data, background
and a Higgs boson signal with $\mh=125$~\GeV.
Final-discriminant bins in all  regions
are combined into bins of $\log_{10}(S/B)$,
with $S$ being the fitted signal and $B$ the fitted background yields.
The Higgs boson signal contribution is shown after rescaling the SM cross-section
according to the value of the signal strength extracted from data ($\mu=1.16$).
In the lower panel, the pull of the data relative to the background (the statistical
significance of the difference between data and fitted background) is shown with
statistical uncertainties only. The full line indicates the pull expected from the sum of
fitted signal and background relative to the fitted background.
}
\label{fig:logstob}
\end{figure}

Table~\ref{tab:results} shows the signal strengths, $p_0$ and significance values from the
combined fit with a single signal strength, and from a fit where the lepton channels each have their own signal strength.
The probability that the signal strengths measured in the three
lepton channels\footnote{The probability of compatibility between fits differing only in their number of
parameters of interest is evaluated in the asymptotics regime, where the difference
between their maximum likelihoods follows a $\chi^{2}$ distribution with a number of
degrees of freedom equal to the difference between the numbers of parameters of
interest.} are compatible is 80\%.
 
{\renewcommand{\arraystretch}{1.3}
\begin{table}[phtb]
\caption{Measured signal strengths with their combined statistical and systematic
uncertainties, expected and observed $p_{0}$ and significance values
(in standard deviations) from the combined fit with a single signal strength, and from a combined fit where each of the lepton channels has its own signal strength, using 13 \TeV\ data.
\label{tab:results}}
\begin{center}
\begin{tabular}{lccccc}
\toprule
\multirow{2}{*}{Signal strength} & \multirow{2}{*}{Signal strength} & \multicolumn{2}{c}{$p_{0}$} & \multicolumn{2}{c}{Significance} \\
\cmidrule(lr){3-4}
\cmidrule(lr){5-6}
& & Exp. & Obs.                  & Exp. & Obs. \\
\midrule
0-lepton     &  $1.04 ^{+0.34}_{-0.32}$   & $9.5\cdot10^{-4}$ & $5.1\cdot10^{-4}$ & $3.1$ & $3.3$   \\
1-lepton     &  $1.09 ^{+0.46}_{-0.42}$   & $8.7\cdot10^{-3}$ & $4.9\cdot10^{-3}$ & $2.4$ & $2.6$\\
2-lepton     &  $1.38 ^{+0.46}_{-0.42}$   & $4.0\cdot10^{-3}$ & $3.3\cdot10^{-4}$ & $2.6$ & $3.4$\\
\midrule
$VH$, $H\to\bbbar$ combination   &  $1.16 ^{+0.27}_{-0.25}$   &  $7.3\cdot10^{-6}$&  $5.3\cdot 10^{-7}$ & $4.3$ & $4.9$ \\
\bottomrule
\end{tabular}
\end{center}
\end{table}
}
 
A combined fit is also performed with floating signal strengths separately for
the $WH$ and $ZH$ production processes. The results of this fit are shown in
Figure~\ref{fig:mu-higgs-c}.
The $WH$ and $ZH$ production modes have observed (expected) significances of 2.5 (2.3) and 4.0 (3.5)
standard deviations, respectively, with a linear correlation between the two signal
strengths of $-1\%$.
 
\begin{figure}
\begin{center}
\includegraphics[width=0.7\textwidth]{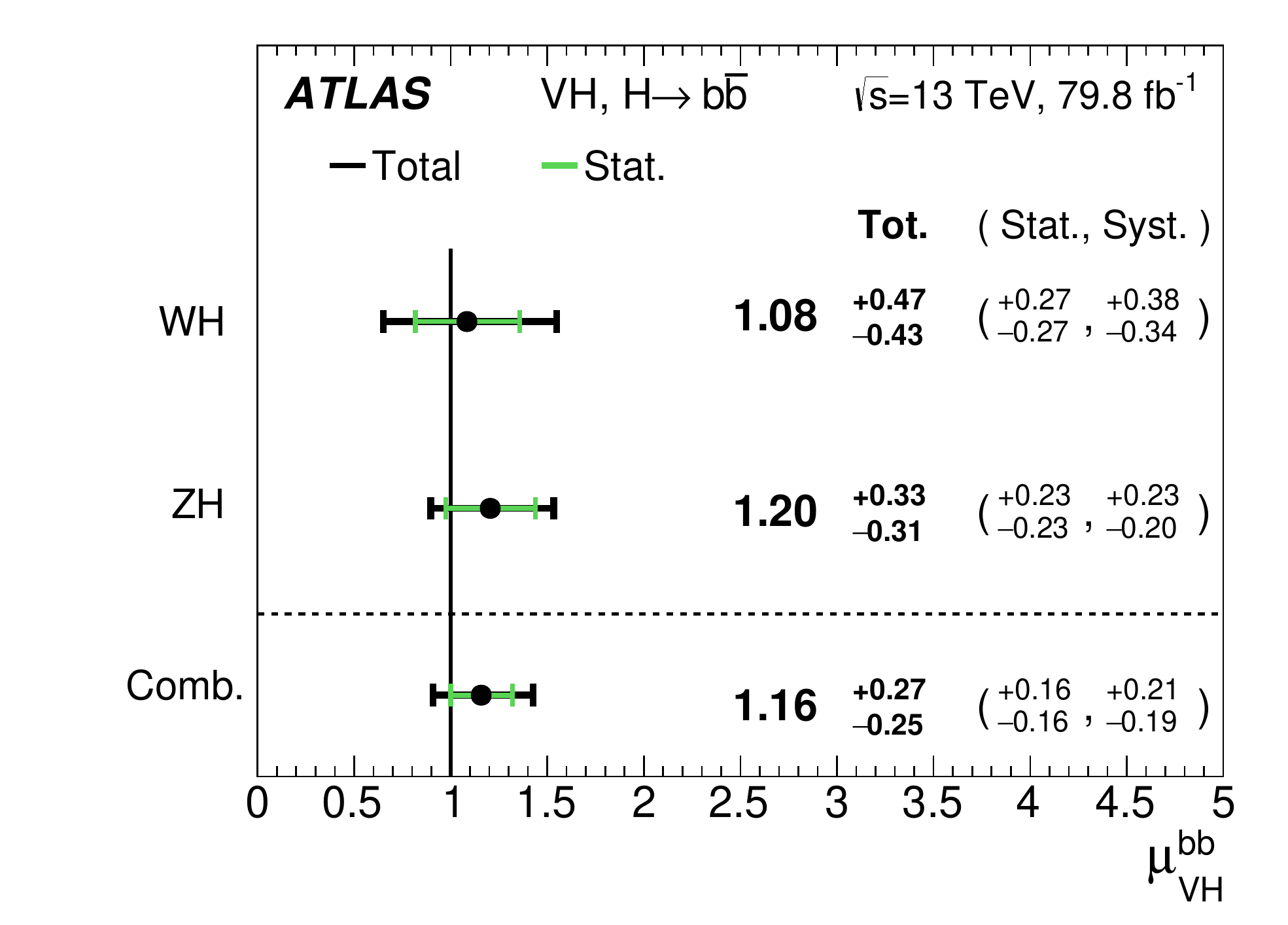}
\end{center}
\caption{The fitted values of the Higgs boson signal strength $\mu_{VH}^{bb}$ for $\mh=125$~\GeV\ for the $WH$ and $ZH$ processes and their combination. The individual $\mu_{VH}^{bb}$ values for the $(W/Z)H$ processes
are obtained from a simultaneous fit with the signal strength for each of the $WH$ and $ZH$ processes floating independently.
The probability of compatibility of the individual signal strengths is 84\%.\label{fig:mu-higgs-c}}
\end{figure}

\FloatBarrier
 
\subsection{Results of the dijet-mass analysis}
 
For all channels combined the fitted value of the signal strength is
\begin{eqnarray*}
\mu_{VH}^{bb} = 1.06^{+0.36}_{-0.33} = 1.06 \pm 0.20 \mathrm{(stat.)} ^{+0.30}_{-0.26} \mathrm{(syst.)},
\end{eqnarray*}
in good agreement with the result of the multivariate analysis.
The observed excess has a significance of 3.6 standard deviations,
compared to an expectation of 3.5 standard deviations.
Good agreement is also found when comparing the values of signal strengths in the individual channels
from the dijet-mass analysis with those from the multivariate analysis.
 
The \mbb\ distribution is shown in Figure~\ref{fig:mbb_cutbased} summed over all channels and
regions, weighted by their respective values of the ratio of fitted Higgs boson
signal and background yields and after subtraction of all backgrounds except for the
$WZ$ and $ZZ$ diboson processes.
 
\begin{figure}[hbp!]
\begin{center}
\includegraphics[width=0.59\textwidth]{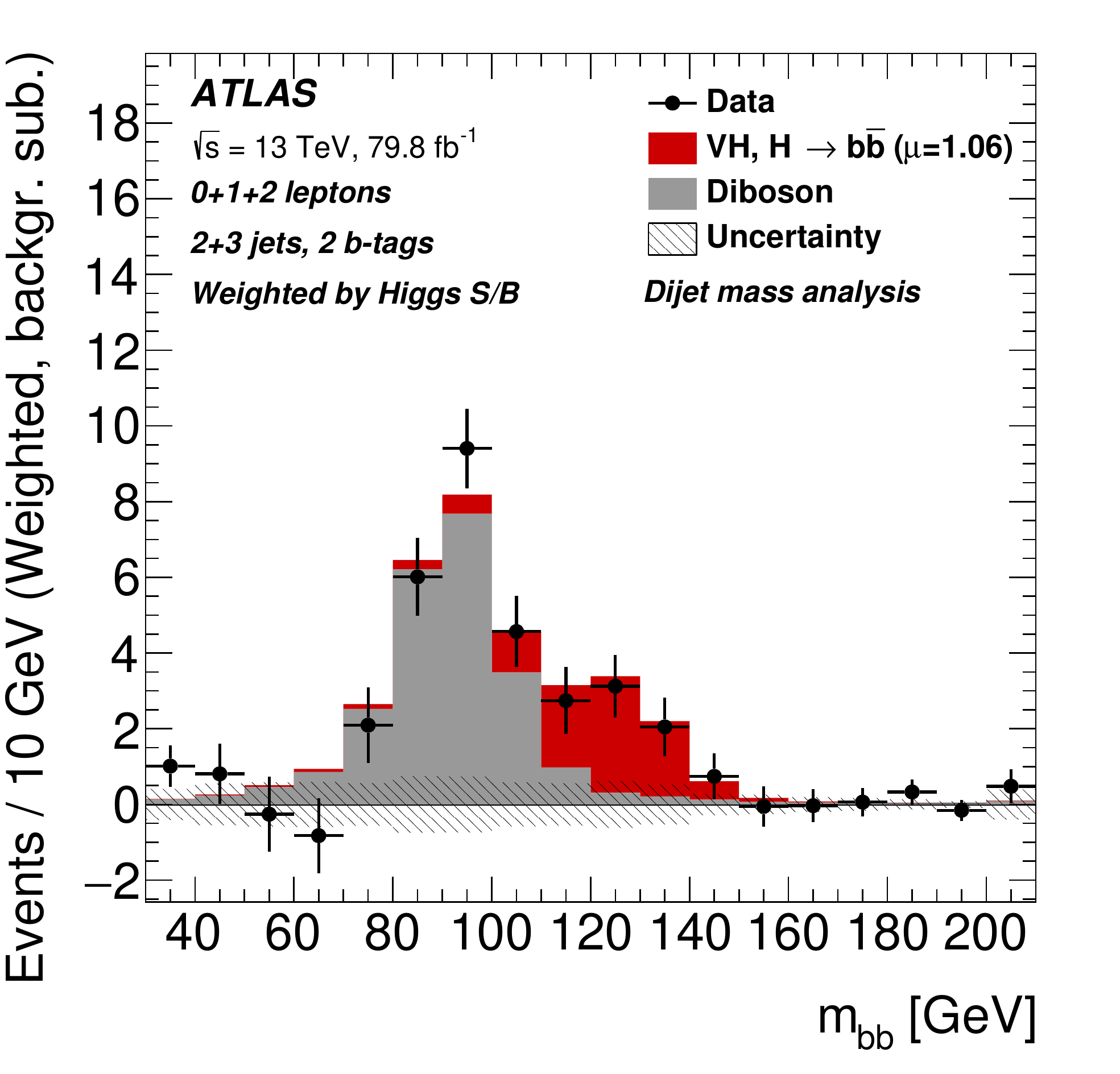}
\end{center}
\caption{The distribution of \mbb\ in data after subtraction of all backgrounds
except for the $WZ$ and $ZZ$ diboson processes,
as obtained with the dijet-mass analysis.
The contributions from all lepton channels, \ptv\ regions and
number-of-jets categories are summed and weighted by their respective $S/B$,
with $S$ being the total fitted signal and $B$ the total fitted background in each region.
The expected contribution of the associated $WH$ and $ZH$ production of a SM Higgs boson with $\mh=125$~\GeV\
is shown scaled by the measured signal strength ($\mu = 1.06$).
The size of the combined statistical and systematic uncertainty
for the fitted background is indicated by the hatched band.
\label{fig:mbb_cutbased}}
\end{figure}
 
\subsection{Results of the diboson analysis}
 
As a validation of the Higgs boson search analysis, the measurement of $VZ$ production based on the multivariate analysis described in Section~\ref{subsec:dibosonfit} returns a value of signal strength
\begin{eqnarray*}
\mu_{VZ}^{bb} = 1.20 ^{+0.20}_{-0.18} = 1.20 \pm 0.08 \mathrm{(stat.)} ^{+ 0.19}_{-0.16} \mathrm{(syst.)},
\end{eqnarray*}
 
in good agreement with the Standard Model prediction.
Analogously to the $VH$ signal, fits are also performed with separate signal strengths for the $WZ$ and $ZZ$ production modes, and the results are shown in Figure~\ref{fig:mu-higgs-c-dib}.
 
\begin{figure}[tbhp!]
\begin{center}
\includegraphics[width=0.7\textwidth]{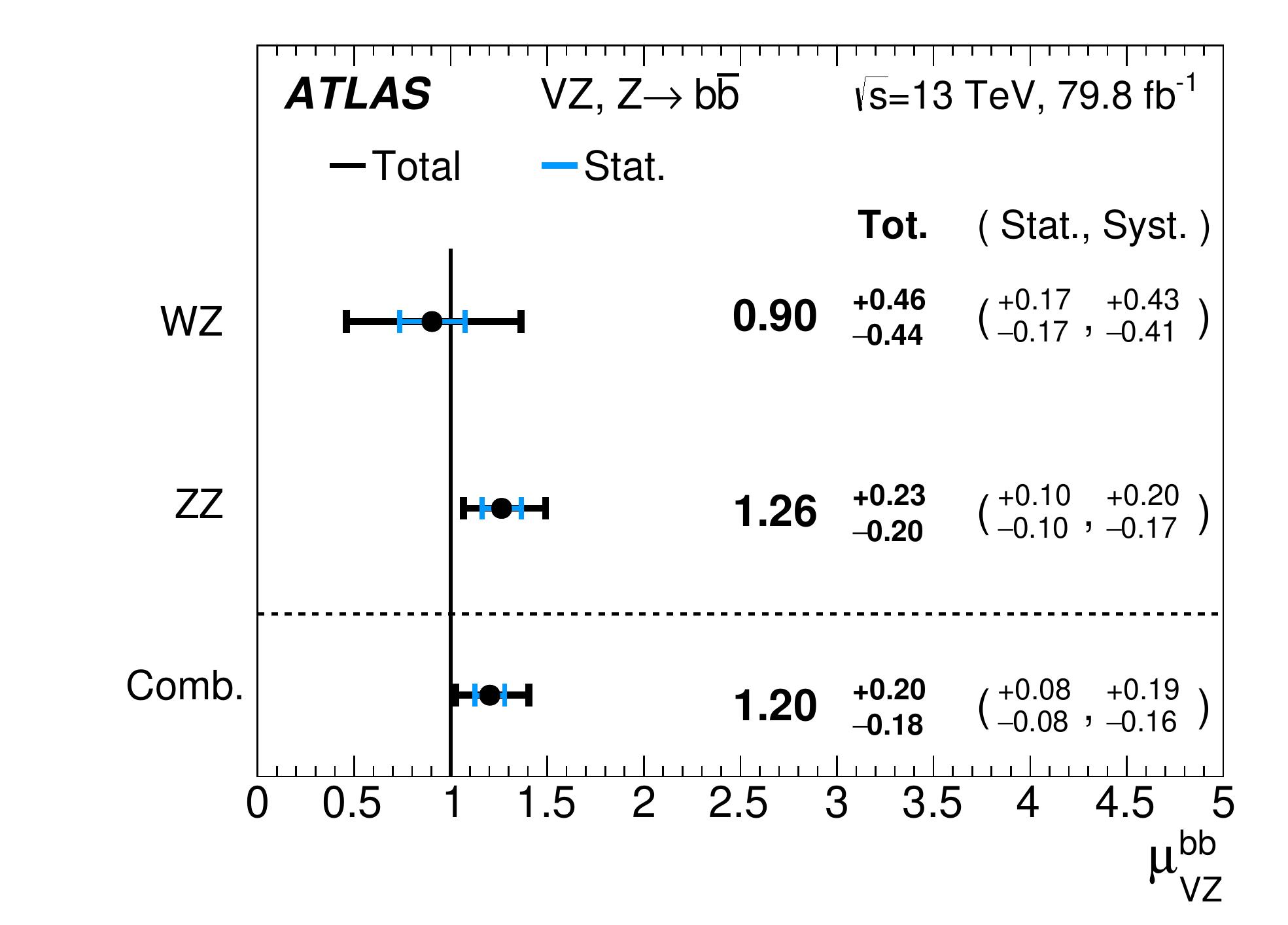}
\end{center}
\caption{The fitted values of the $VZ$  signal strength $\mu_{VZ}^{bb}$ for the $WZ$ and $ZZ$ processes
and their combination. The individual $\mu_{VZ}^{bb}$ values for the $(W/Z)Z$ processes
are obtained from a simultaneous fit with the signal strengths for each of the $WZ$ and $ZZ$ processes floating independently.
The probability of compatibility of the individual signal strengths is 47\%.
\label{fig:mu-higgs-c-dib}}
\end{figure}
 
\FloatBarrier
 
\subsection{Results of combinations}
 
\subsubsection{Run 1 and Run 2 combination for $VH$, $H\to\bbbar$}
 
The result of the Run~2 analysis is combined with the Run~1 $VH$, $H\to\bbbar$ result following the methodology
described in Section~\ref{subsec:fitcombination}. The observed $p_0$ value is $5.5\cdot10^{-7}$, corresponding to
an excess with a significance of 4.9 standard deviations, compared with an expectation of
5.1 standard deviations. The measured signal strength is:
\begin{eqnarray*}
\mu^{bb}_{VH} = 0.98 ^{+0.22}_{-0.21} = 0.98 \pm 0.14 \mathrm{(stat.)} ^{+0.17}_{-0.16} \mathrm{(syst.)}.
\end{eqnarray*}
Fits are also performed with the signal strengths floated independently for the
$WH$ and $ZH$ production processes. The probability of compatibility of the signal strengths for the $WH$
and $ZH$ production processes is 72\%, and the results of this fit are shown in Figure~\ref{fig:mu-higgs-c-comb}.
 
\begin{figure}[htb!p]
\begin{center}
\includegraphics[width=0.7\textwidth]{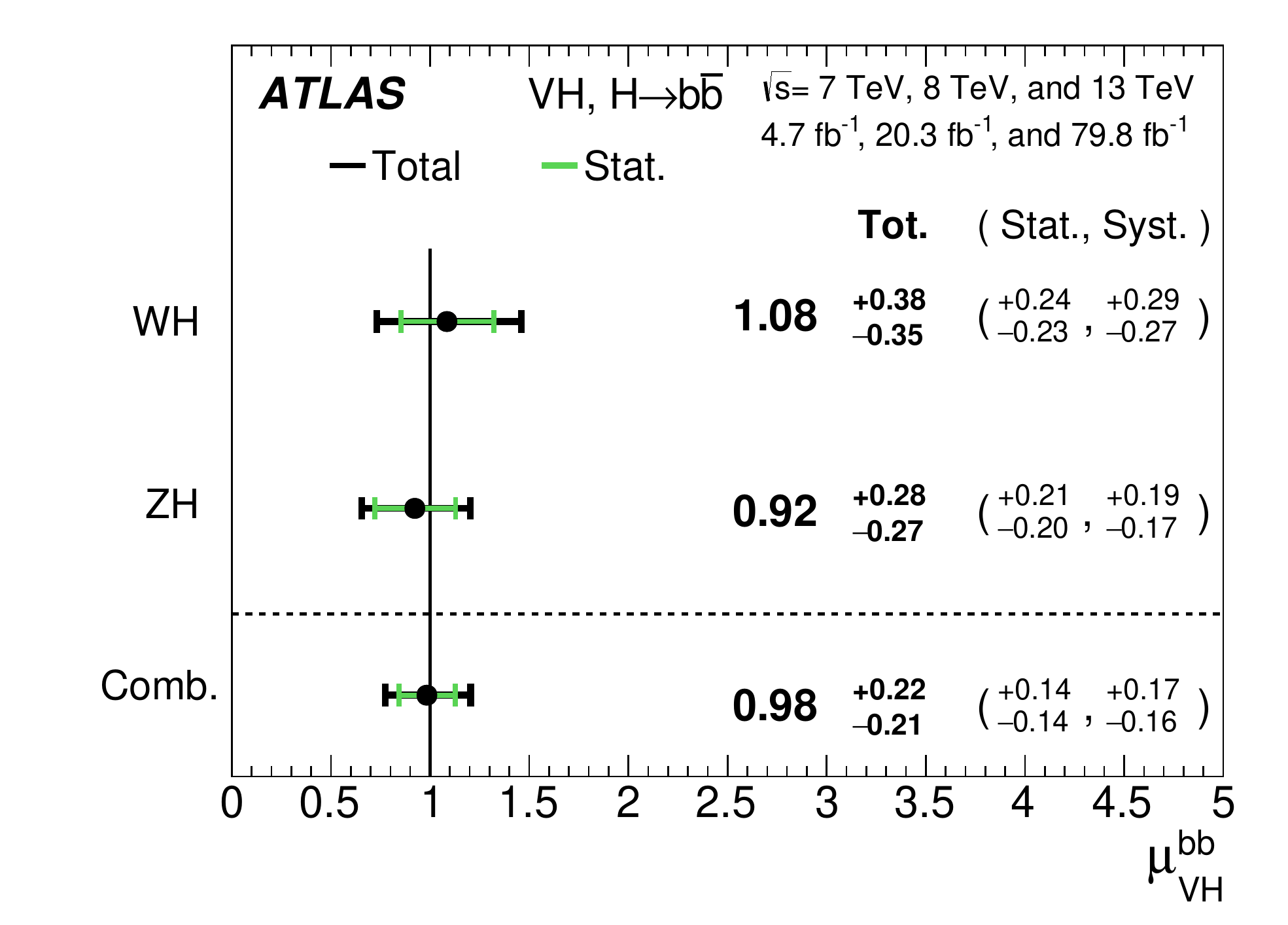}
\end{center}
\caption{The fitted values of the Higgs boson signal strength $\mu_{VH}^{bb}$ for
$\mh=125$~\GeV\ for the $WH$ and $ZH$ processes and their combination, using the 7\,\TeV, 8\,\TeV\ and 13\,\TeV\ data. The individual $\mu_{VH}^{bb}$ values for the $(W/Z)H$
processes are obtained from a simultaneous fit with the signal strengths for
each of the $WH$ and $ZH$ processes floating independently.\label{fig:mu-higgs-c-comb}}
\end{figure}
 
\subsubsection{Observation of $H\to\bbbar$ decays}
 
The $VH$ result is further combined  with results of the searches for the Standard Model Higgs boson
decaying into a $b\bar{b}$ pair produced in association with a $t\bar{t}$ pair and in
vector-boson fusion for both Run~1 and Run~2, to perform a search for the
$H\to\bbbar$ decay. For a Higgs boson mass of 125 \GeV, and assuming the relative
production cross-sections are those predicted by the SM, the observed significance for the $H\to\bbbar$ decay is 5.4 standard deviations,
to be compared with an expectation of 5.5 standard deviations. With the additional
assumption that the production cross-sections are those predicted by the SM, the fitted value for all channels combined of the
signal strength of the branching fraction into $b$-quark pairs is
\begin{eqnarray*}
\mu_{H\to bb} = 1.01 \pm 0.20 = 1.01 \pm 0.12 \mathrm{(stat.)} ^{+0.16}_{-0.15} \mathrm{(syst.)}.
\end{eqnarray*}
Table~\ref{tab:mu-higgs-b-comb-results} shows the significance values independently for
the VBF+ggF, $\ttbar H$ and $VH$ channels in the combination of the Run~1 and Run~2 data, and
for the combined global likelihood fit. The signal strengths obtained from a fit where individual
signal strengths are fitted simultaneously for the three production modes are displayed in
Figure~\ref{fig:mu-higgs-b-comb}. Fits are also performed with the signal strengths
floated  independently for each of the production processes in both Run~1 and
Run~2. The probability of compatibility of the six individual measurements is 54\%.
 
\begin{table}[phtb]
\caption{Expected and observed significance values (in
standard deviations) for the $H \to b\bar{b}$ channels fitted independently and their combination using the 7  \TeV, 8 \TeV\ and 13 \TeV\ data.
\label{tab:mu-higgs-b-comb-results}}
\begin{center}
\begin{tabular}{lcc}
\toprule
\multirow{2}{*}{Channel} &  \multicolumn{2}{c}{Significance} \\
\cmidrule(lr){2-3}
& Exp. & Obs.    \\
\midrule
VBF+ggF              & $0.9$ & $1.5$   \\
$\ttbar H$   & $1.9$ & $1.9$ \\
$VH$           & $5.1$ & $4.9$ \\
\midrule
$H\to \bbbar$ combination     & $5.5$ & $5.4$ \\
\bottomrule
\end{tabular}
\end{center}
\end{table}
 
\begin{figure}[htb!p]
\begin{center}
\includegraphics[width=0.7\textwidth]{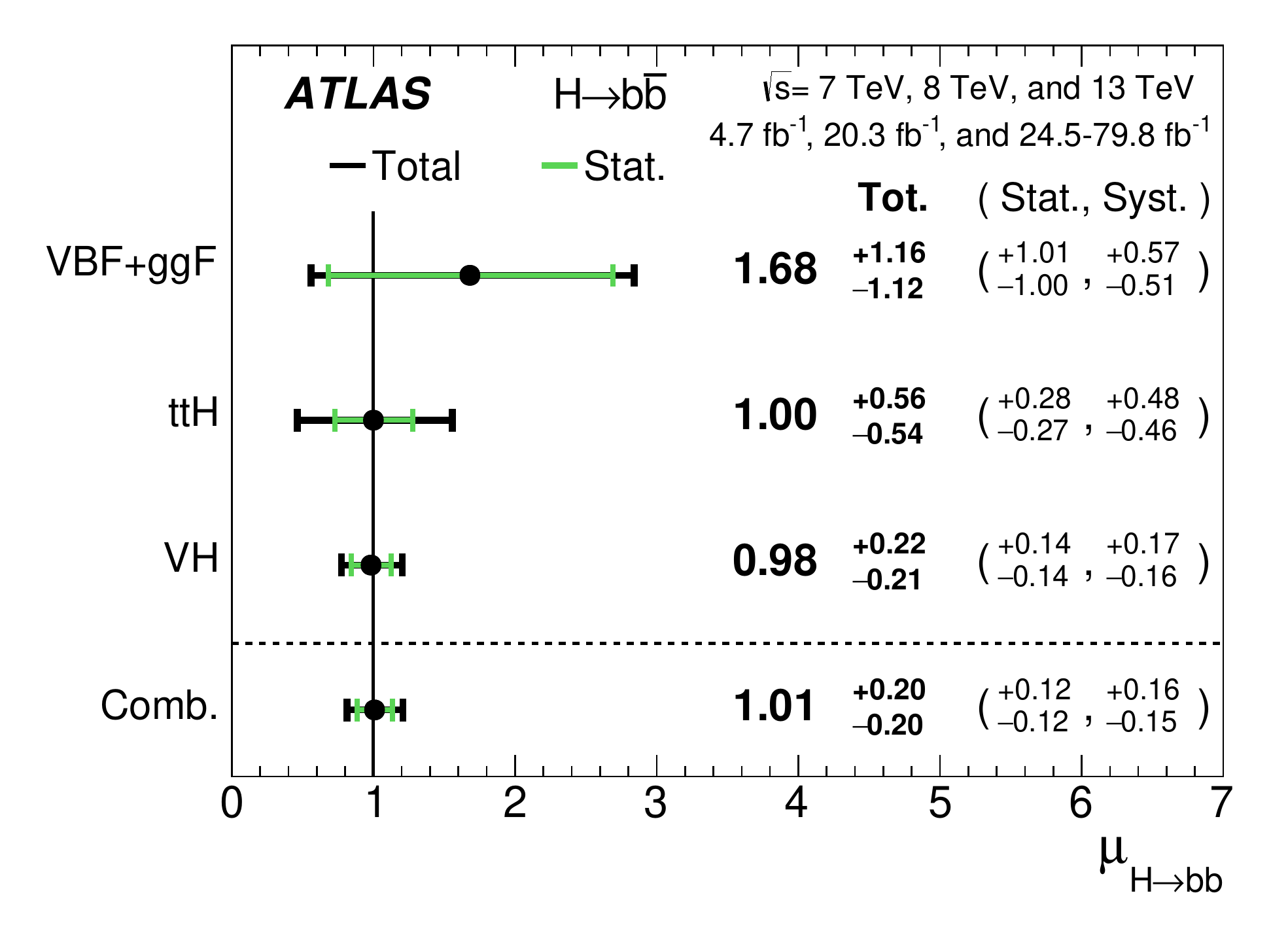}
\end{center}
\caption{The fitted values of the Higgs boson signal strength $\mu_{H \to bb}$ for
$\mh=125$~\GeV\ separately for the $VH$, $t\bar{t}H$ and VBF+ggF analyses along
with their combination, using the 7\,\TeV, 8\,\TeV\ and 13\,\TeV\ data.
The individual $\mu_{H \to bb}$ values for the different
production modes are obtained from a simultaneous fit with the signal strengths
for each of the processes floating independently. The probability of compatibility of the
individual signal strengths is 83\%.
\label{fig:mu-higgs-b-comb}}
\end{figure}
 
\subsubsection{Observation of $VH$ production}
 
The Run~2 $VH$, $H\to b\bar{b}$ result is further combined with the results of other Run~2 searches for
the Higgs boson produced in the $VH$ production mode, but decaying into either two photons or
four leptons via $ZZ^*$ decays. For a Higgs boson mass of 125~\GeV, and assuming the relative branching fractions of the three decay modes considered to be as predicted by the SM, the observed
significance for $VH$ production is 5.3 standard deviations, to be compared with an expectation of 4.8 standard deviations.
Table~\ref{tab:mu-higgs-VH-comb-results} shows the significance values for the
combined global likelihood fit, and for a fit where the four-lepton ($H \to ZZ^* \to
4\ell$), diphoton ($H \to \gamma\gamma$) and $H \to b\bar{b}$ decay modes each have their own signal strength for the Run~2 data.
Assuming the branching fractions are as predicted by the SM, the fitted value of the $VH$ signal strength for all channels
combined is:
\begin{eqnarray*}
\mu_{VH} = 1.13 ^{+0.24}_{-0.23} = 1.13 \pm 0.15 \mathrm{(stat.)} ^{+0.18}_{-0.17} \mathrm{(syst.)}.
\end{eqnarray*}
The signal strengths obtained from the fit where individual
signal strengths are fitted for the three decay modes are displayed in
Figure~\ref{fig:mu-vh-comb}, along with their combination.
 
\begin{figure}[htb!p]
\begin{center}
\includegraphics[width=0.7\textwidth]{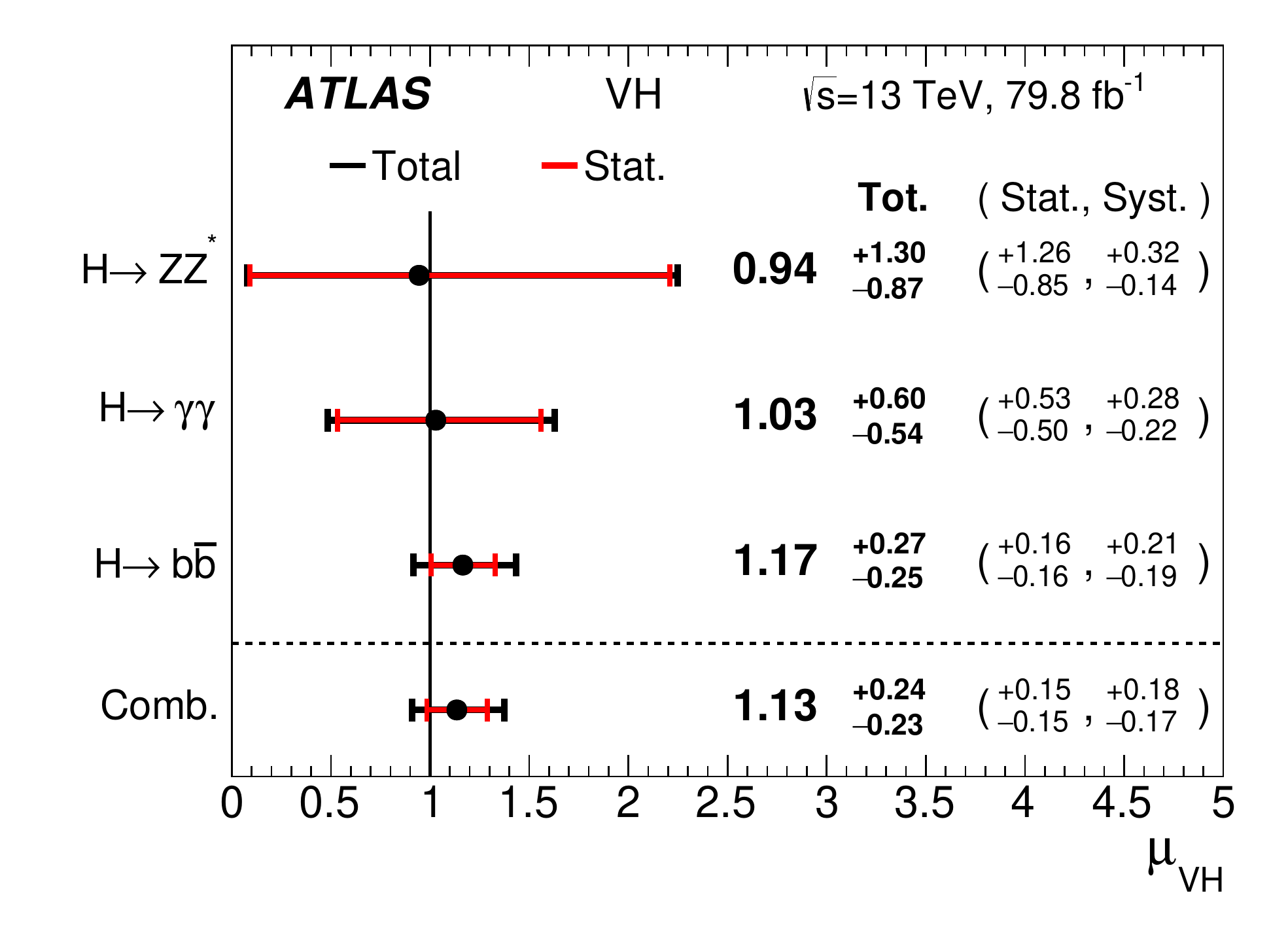}
\end{center}
\caption{The fitted values of the Higgs boson signal strength $\mu_{VH}$ for
$\mh=125$~\GeV\ separately for the $H \to b\bar{b}$, $H \to \gamma \gamma $ and $H \to
ZZ^{*} \to 4\ell$ decay modes, along
with their combination. The individual $\mu_{VH}$ values for the different
decay modes are obtained from a simultaneous fit with the signal strengths
for each of the processes floating independently. The probability of compatibility of the
individual signal strengths is 96\%.
\label{fig:mu-vh-comb}}
\end{figure}
 
\begin{table}[t!phb]
\caption{Expected and observed significance values (in
standard deviations) for the $VH$ production channels from the combined fit and from a combined fit where each of the lepton channels has its own signal strength, using 13~\TeV\ data.
\label{tab:mu-higgs-VH-comb-results}}
\begin{center}
\begin{tabular}{lcc}
\toprule
\multirow{2}{*}{Channel} &  \multicolumn{2}{c}{Significance} \\
\cmidrule(lr){2-3}
& Exp. & Obs.    \\
\midrule
$H \to ZZ^* \to 4\ell$              & $1.1$ & $1.1$   \\
$H \to\gamma\gamma$        & $1.9$ & $1.9$ \\
$H \to b\bar{b}$                     & $4.3$ & $4.9$ \\
\midrule
VH combined     & $4.8$ & $5.3$ \\
\bottomrule
\end{tabular}
\end{center}
\end{table}
 
% End of text imported from the .//paper_08_results.tex input file

\FloatBarrier

\section{Conclusion}
% The next lines are included from the .//paper_09_conclusions.tex input file
A search for the Standard Model Higgs boson decaying into a $b\bar{b}$ pair and produced in
association with a $W$ or $Z$ boson is presented, using data collected by the ATLAS
experiment in proton--proton collisions from Run~2 of the LHC\@. The data
correspond to an integrated luminosity of \lumiFull\ collected at a centre-of-mass energy of $\sqrt{s} = $13~\TeV.
An excess over the expected background is observed, with a significance of 4.9 standard
deviations compared with an expectation of 4.3. The measured signal strength relative
to the SM prediction for $m_H = 125$~\GeV\ is found to be
$\mu_{VH}^{bb}~=~1.16 \pm 0.16 \mathrm{(stat.)} ^{+0.21}_{-0.19} \mathrm{(syst.)}$.
 
This result is combined with previous results based on all the Run~1 data collected at
centre-of-mass energies of 7~\TeV\ and 8~\TeV. An excess over the expected SM
background is observed, with a significance of 4.9 standard deviations compared with an
expectation of 5.1. The measured signal strength relative to the SM  expectation is found to be
$\mu_{VH}^{bb} = 0.98 \pm 0.14 \mathrm{(stat.)} ^{+0.17}_{-0.16} \mathrm{(syst.)}$.
 
Results for the SM Higgs boson decaying into a $b\bar{b}$ pair in the $VH$,
$\ttbar H$ and VBF+ggF production modes at centre-of-mass energies of 7~\TeV, 8~\TeV\ and
13~\TeV\ are also combined, assuming the relative production cross-sections of these
processes to be as predicted by the SM\@. An excess over the expected SM background is
observed, with a significance of 5.4 standard deviations compared with
an expectation of 5.5, providing an observation of the $H \to b\bar{b}$ decay mode. Assuming
the SM production strengths, the measured signal strength relative to the
SM expectation is $\mu_{H \to
bb} = 1.01 \pm 0.12 \mathrm{(stat.)} ^{+0.16}_{-0.15} \mathrm{(syst.)}$, consistent with the value of the
Yukawa coupling to bottom quarks in the SM\@.
 
In addition, the Run~2 $VH$, $H\to b\bar{b}$ result is further combined with the results
of other Run~2
searches for the Higgs boson decaying into either four leptons (via $ZZ^*$) or diphotons in
the $VH$ production mode, assuming the relative branching fractions of
the three decay modes to be as predicted by the SM\@. The result is an observed
significance of 5.3 standard deviations, to be compared with an expectation of 4.8 standard
deviations. Assuming the SM branching fractions, the measured signal strength
relative to the SM expectation is
$ \mu_{VH} = 1.13 \pm 0.15 \mathrm{(stat.)} ^{+0.18}_{-0.17} \mathrm{(syst.)}$.
This provides a direct observation of the Higgs boson being produced in
association with a vector boson.
 
% End of text imported from the .//paper_09_conclusions.tex input file

\section*{Acknowledgements}

We thank CERN for the very successful operation of the LHC, as well as the
support staff from our institutions without whom ATLAS could not be
operated efficiently.
 
We acknowledge the support of ANPCyT, Argentina; YerPhI, Armenia; ARC, Australia; BMWFW and FWF, Austria; ANAS, Azerbaijan; SSTC, Belarus; CNPq and FAPESP, Brazil; NSERC, NRC and CFI, Canada; CERN; CONICYT, Chile; CAS, MOST and NSFC, China; COLCIENCIAS, Colombia; MSMT CR, MPO CR and VSC CR, Czech Republic; DNRF and DNSRC, Denmark; IN2P3-CNRS, CEA-DRF/IRFU, France; SRNSFG, Georgia; BMBF, HGF, and MPG, Germany; GSRT, Greece; RGC, Hong Kong SAR, China; ISF, I-CORE and Benoziyo Center, Israel; INFN, Italy; MEXT and JSPS, Japan; CNRST, Morocco; NWO, Netherlands; RCN, Norway; MNiSW and NCN, Poland; FCT, Portugal; MNE/IFA, Romania; MES of Russia and NRC KI, Russian Federation; JINR; MESTD, Serbia; MSSR, Slovakia; ARRS and MIZ\v{S}, Slovenia; DST/NRF, South Africa; MINECO, Spain; SRC and Wallenberg Foundation, Sweden; SERI, SNSF and Cantons of Bern and Geneva, Switzerland; MOST, Taiwan; TAEK, Turkey; STFC, United Kingdom; DOE and NSF, United States of America. In addition, individual groups and members have received support from BCKDF, the Canada Council, CANARIE, CRC, Compute Canada, FQRNT, and the Ontario Innovation Trust, Canada; EPLANET, ERC, ERDF, FP7, Horizon 2020 and Marie Sk{\l}odowska-Curie Actions, European Union; Investissements d'Avenir Labex and Idex, ANR, R{\'e}gion Auvergne and Fondation Partager le Savoir, France; DFG and AvH Foundation, Germany; Herakleitos, Thales and Aristeia programmes co-financed by EU-ESF and the Greek NSRF; BSF, GIF and Minerva, Israel; BRF, Norway; CERCA Programme Generalitat de Catalunya, Generalitat Valenciana, Spain; the Royal Society and Leverhulme Trust, United Kingdom.
 
The crucial computing support from all WLCG partners is acknowledged gratefully, in particular from CERN, the ATLAS Tier-1 facilities at TRIUMF (Canada), NDGF (Denmark, Norway, Sweden), CC-IN2P3 (France), KIT/GridKA (Germany), INFN-CNAF (Italy), NL-T1 (Netherlands), PIC (Spain), ASGC (Taiwan), RAL (UK) and BNL (USA), the Tier-2 facilities worldwide and large non-WLCG resource providers. Major contributors of computing resources are listed in Ref.~\cite{ATL-GEN-PUB-2016-002}.

\clearpage
 
\printbibliography
 
\clearpage
 
% ATLAS Collaboration author list
% Reference date of HIGG-2018-04 is 2018-07-13
% Author list last updated on date 16-AUG-18
% Data extracted on 16-Aug-2018 for paper reference HIGG-2018-04
% at 12:27pm
 
\begin{flushleft}
{\Large The ATLAS Collaboration}

\bigskip

M.~Aaboud$^\textrm{\scriptsize 34d}$,    
G.~Aad$^\textrm{\scriptsize 99}$,    
B.~Abbott$^\textrm{\scriptsize 124}$,    
O.~Abdinov$^\textrm{\scriptsize 13,*}$,    
B.~Abeloos$^\textrm{\scriptsize 128}$,    
D.K.~Abhayasinghe$^\textrm{\scriptsize 91}$,    
S.H.~Abidi$^\textrm{\scriptsize 164}$,    
O.S.~AbouZeid$^\textrm{\scriptsize 39}$,    
N.L.~Abraham$^\textrm{\scriptsize 153}$,    
H.~Abramowicz$^\textrm{\scriptsize 158}$
% ...
\bigskip
\\

% ...
$^{13}$Institute of Physics, Azerbaijan Academy of Sciences, Baku; Azerbaijan.\\
% ...
$^{34}$$^{(a)}$Facult\'e des Sciences Ain Chock, R\'eseau Universitaire de Physique des Hautes Energies - Universit\'e Hassan II, Casablanca;$^{(b)}$Centre National de l'Energie des Sciences Techniques Nucleaires (CNESTEN), Rabat;$^{(c)}$Facult\'e des Sciences Semlalia, Universit\'e Cadi Ayyad, LPHEA-Marrakech;$^{(d)}$Facult\'e des Sciences, Universit\'e Mohamed Premier and LPTPM, Oujda;$^{(e)}$Facult\'e des sciences, Universit\'e Mohammed V, Rabat; Morocco.\\
% ...
$^{39}$Niels Bohr Institute, University of Copenhagen, Copenhagen; Denmark.\\
% ...
$^{91}$Department of Physics, Royal Holloway University of London, Egham; United Kingdom.\\
% ...
$^{99}$CPPM, Aix-Marseille Universit\'e, CNRS/IN2P3, Marseille; France.\\
% ...
$^{124}$Homer L. Dodge Department of Physics and Astronomy, University of Oklahoma, Norman OK; United States of America.\\
% ...
$^{128}$LAL, Universit\'e Paris-Sud, CNRS/IN2P3, Universit\'e Paris-Saclay, Orsay; France.\\
% ...
$^{153}$Department of Physics and Astronomy, University of Sussex, Brighton; United Kingdom.\\
% ...
$^{158}$Raymond and Beverly Sackler School of Physics and Astronomy, Tel Aviv University, Tel Aviv; Israel.\\
% ...
$^{164}$Department of Physics, University of Toronto, Toronto ON; Canada.\\
% ...
$^{a}$ Also at Borough of Manhattan Community College, City University of New York, NY; United States of America.\\
$^{b}$ Also at California State University, East Bay; United States of America.\\
$^{c}$ Also at Centre for High Performance Computing, CSIR Campus, Rosebank, Cape Town; South Africa.\\
$^{d}$ Also at CERN, Geneva; Switzerland.\\
$^{e}$ Also at CPPM, Aix-Marseille Universit\'e, CNRS/IN2P3, Marseille; France.\\
% ...
$^{*}$ Deceased

\end{flushleft}

% Created with Glance <Atlas.Glance@cern.ch>

\end{document}